\documentclass[10pt,letterpaper]{article}
\usepackage[top=0.85in,left=2.75in,footskip=0.75in]{geometry}

\usepackage{graphicx}
\usepackage{gensymb}
\usepackage{textcomp}

\usepackage{amsmath,amssymb}

\usepackage{changepage}

\usepackage[utf8x]{inputenc}

\usepackage{textcomp,marvosym}

\usepackage{cite}

\usepackage{nameref,hyperref}

\usepackage[right]{lineno}

\usepackage{microtype}
\DisableLigatures[f]{encoding = *, family = * }

\usepackage[table]{xcolor}

\usepackage{array}

\newcolumntype{+}{!{\vrule width 2pt}}

\newlength\savedwidth

\raggedright
\usepackage[aboveskip=1pt,labelfont=bf,labelsep=period,justification=raggedright,singlelinecheck=off]{caption}

\makeatletter
\renewcommand{\@biblabel}[1]{\quad#1.}
\makeatother

\date{}

\usepackage{lastpage,fancyhdr,graphicx}
\usepackage{epstopdf}
\fancyheadoffset[L]{2.25in}
\fancyfootoffset[L]{2.25in}

\begin{document}
\vspace*{0.2in}

\begin{flushleft}
{\Large
\textbf\newline{State Dependence of Stimulus-Induced Variability Tuning in Macaque MT} 
}
\newline
\\  
Joseph A.\ Lombardo\textsuperscript{1,2},
Matthew V.\ Macellaio\textsuperscript{3,4},
Bing Liu\textsuperscript{4},
Stephanie E.\ Palmer\textsuperscript{2,5\Yinyang*},
with Leslie C.\ Osborne\textsuperscript{2,4,6\Yinyang*}
\\
\bigskip
\textbf{1} Computational Neuroscience Graduate Program, University of Chicago, Chicago, IL USA
\\
\textbf{2} Department of Organismal Biology and Anatomy, University of Chicago, Chicago, IL USA
\\
\textbf{3} Neurobiology Graduate Program, University of Chicago, Chicago, IL USA
\\
\textbf{4} Department of Neurobiology, University of Chicago, Chicago, IL USA
\\
\textbf{5} Department of Physics, University of Chicago, Chicago, IL USA
\\
\textbf{6} Present address: Department of Neurobiology, Duke University, Durham, NC USA
\bigskip

%
%
\Yinyang These authors contributed equally to this work.





* sepalmer@uchicago.edu, osborne.leslie@gmail.com

\end{flushleft}
\section*{Abstract}

Behavioral states marked by varying levels of arousal and attention modulate some properties of cortical responses (e.g.\ average firing rates or pairwise correlations), yet it is not fully understood what drives these response changes and how they might affect downstream stimulus decoding. Here we show that changes in state modulate the tuning of response variance-to-mean ratios (Fano factors) in a fashion that is neither predicted by a Poisson spiking model nor changes in the mean firing rate, with a substantial effect on stimulus discriminability. We recorded motion-sensitive neurons in middle temporal cortex (MT) in two states: alert fixation and light, opioid anesthesia. Anesthesia tended to lower average spike counts, without decreasing trial-to-trial variability compared to the alert state. Under anesthesia, within-trial fluctuations in excitability were correlated over longer time scales compared to the alert state, creating supra-Poisson Fano factors. In contrast, alert-state MT neurons have higher mean firing rates and largely sub-Poisson variability that is stimulus-dependent and cannot be explained by firing rate differences alone. The absence of such stimulus-induced variability tuning in the anesthetized state suggests different sources of variability between states. A simple model explains state-dependent shifts in the distribution of observed Fano factors via a suppression in the variance of gain fluctuations in the alert state. A population model with stimulus-induced variability tuning and behaviorally constrained information-limiting correlations explores the potential enhancement in stimulus discriminability by the cortical population in the alert state.

\section*{Author Summary}
 The brain controls behavior fluidly in a wide variety of cognitive contexts that alter the precision of neural responses. We examine how neural variability changes versus the mean response as a function of the stimulus and the behavioral state. We show that this scaled variability can have qualitatively different stimulus tuning in different behavioral contexts. In alert primates, scaled variability is tuned to the direction of motion of a visual stimulus and decreases around the preferred direction of each neuron. Under anesthesia, neurons show flat scaled variability tuning and, overall, responses are significantly more variable. We develop a simple model that includes a parameter describing firing rate gain fluctuations that can explain these changes. Our results suggest that tuned decreases in scaled variability during wakefulness may be mediated by an active process that suppresses synchronization and makes information transmission more reliable.

\newcommand{\FForth}{\mathrm{FF}_{\mathrm{orth}}}
\newcommand{\FFpref}{\mathrm{FF}_{\mathrm{pref}}}

\section*{Introduction}
Sensory systems operate in many states (e.g.\ attentional states and stages of sleep) wherein the same anatomical network displays different scales of firing rates, variability, correlations, oscillation frequencies, and so forth while maintaining basic function \cite{treue_1999,mcadams_1999,cohen_2009,ecker_state_2014,white_2012,constantinople_bruno_11}. In some states, such as under light anesthesia, cortical networks encode roughly the same information about a stimulus, but with different dynamics and firing rates \cite{brown_2010,chen_2011}. Each neuron's contribution to the accuracy of stimulus decoding depends on its tuning function, the smooth modulation of the firing rate in response to parametric changes in stimulus value, and a neuron's response reliability determines its impact on decoding precision \cite{brunel_nadal_1998,shadlen_variable_1998,montijn_population_2014}. Much is made of how a population code might be either robust or sensitive to neuronal variability, but an important aspect of variability is often left out of the discussion: how does neuronal variability, itself, depend on the stimulus? This question becomes particularly important in the context of how response reliability impacts information transmission over the brain's natural operating range\cite{brunel_nadal_1998,osborne_time_2004,butts_2006,josic_2009} and decoding \cite{brunel_nadal_1998,ma_2006}. 

The impacts of variability on population decoding can be analytically derived for the case of an independent or correlated population of Poisson neurons \cite{brunel_nadal_1998,zhang_1999}, where the mean and variance of the response are the same, yielding a Fano factor $= 1$. The Poisson model of spike generation replicates many of the features of cortical spike trains recorded under some conditions \cite{shadlen_variable_1998}, but the Fano factor prediction is often violated in real sensory neurons \cite{softky_1993,kara_2000,arabzadeh_2004,carandini_2004, montgomery_2010}, particularly when measuring responses over the $\sim 100$ ms timescale of sensory estimation \cite{osborne_time_2004}. Decreases in Fano factor are observed at the onset of visual stimulation, alongside decreases in neuronal correlation, in MT and throughout cortex\cite{osborne_time_2004,smith_2008,cohen_2009,mitchell_2009,churchland_stimulus_2010,white_2012}. 
Moreover, Fano factors can have their own stimulus tuning, which can impact stimulus encoding at the population level \cite{priebe_2012,sadagopan_2012,ponce-alvarez_stimulus-dependent_2013}.

The response properties of neurons in the middle temporal cortical area (MT) have been particularly well described across a number of behavioral states. MT neurons respond selectively to visual motion and firing rates decrease with a Gaussian profile with angular distance from a preferred motion direction \cite{maunsell_functional_1983, albright_direction_1984}. Levels of arousal (anesthesia, alert behavior) and modulations of spatial attention affect the stimulus-averaged excitability of MT neurons, as elsewhere in the brain, but do not tend to shift preferred directions or tuning bandwidths \cite{treue_1999,woolley_11,cook_maunsell_02,mcadams_1999}, much like changing contrast modulates the rate without changing tuning in primary visual cortex \cite{tolhurst_73,dean_81,albrecht_hamilton_82,sclar_freeman_82}. Less well studied are the effects of behavioral state on the variability of cortical responses, a critical measurement for assessing sensory discrimination. Increased attention tends to reduce variability in visual cortical areas, including MT, particularly in narrow-spiking neurons \cite{mitchell_2007,cohen_2009,schledde_2017,galashan_2013,niebergall_2011}. Anesthetic effects may be analogous to a large reduction in attention, decreasing mean firing rates and increasing response variance \cite{brown_annrev_11,white_2012}. 

Here, we explore the state dependence of variability in neural responses and its implications for sensory discriminability. We record single unit responses to motion in MT in alert monkeys and under a light opioid anesthetic as a proxy for a range of natural brain states. Our goal is not to provide an exhaustive review of the myriad effects of anesthesia and other modulations of brain state on cortical variability, but rather to quantify several features of cortical responses under two particular brain states and ask whether a parsimonious model can explain these broad transitions. We find that scaled variability in spike count (Fano factor) is tuned for motion direction, but only in some network states. Alert responses display sub-Poisson variability that is inversely tuned to the stimulus, decreasing at the preferred direction of the cell. Anesthetized responses show flat, supra-Poisson tuning. We identify a simple model through which a single parameter accounts for changes in visually-driven spiking variability in both alert and anesthetized animals. Modulation of the size of gain fluctuations in the response can give rise to two qualitatively different regimes of Fano factor tuning. Finally, we explore how changes in the tuning of the Fano factor influence stimulus discriminability when animals are alert and actively engaged in visual behavior. 

\section*{Results}
\begin{figure}[ht]
	\centering
		\includegraphics[width=1\textwidth]{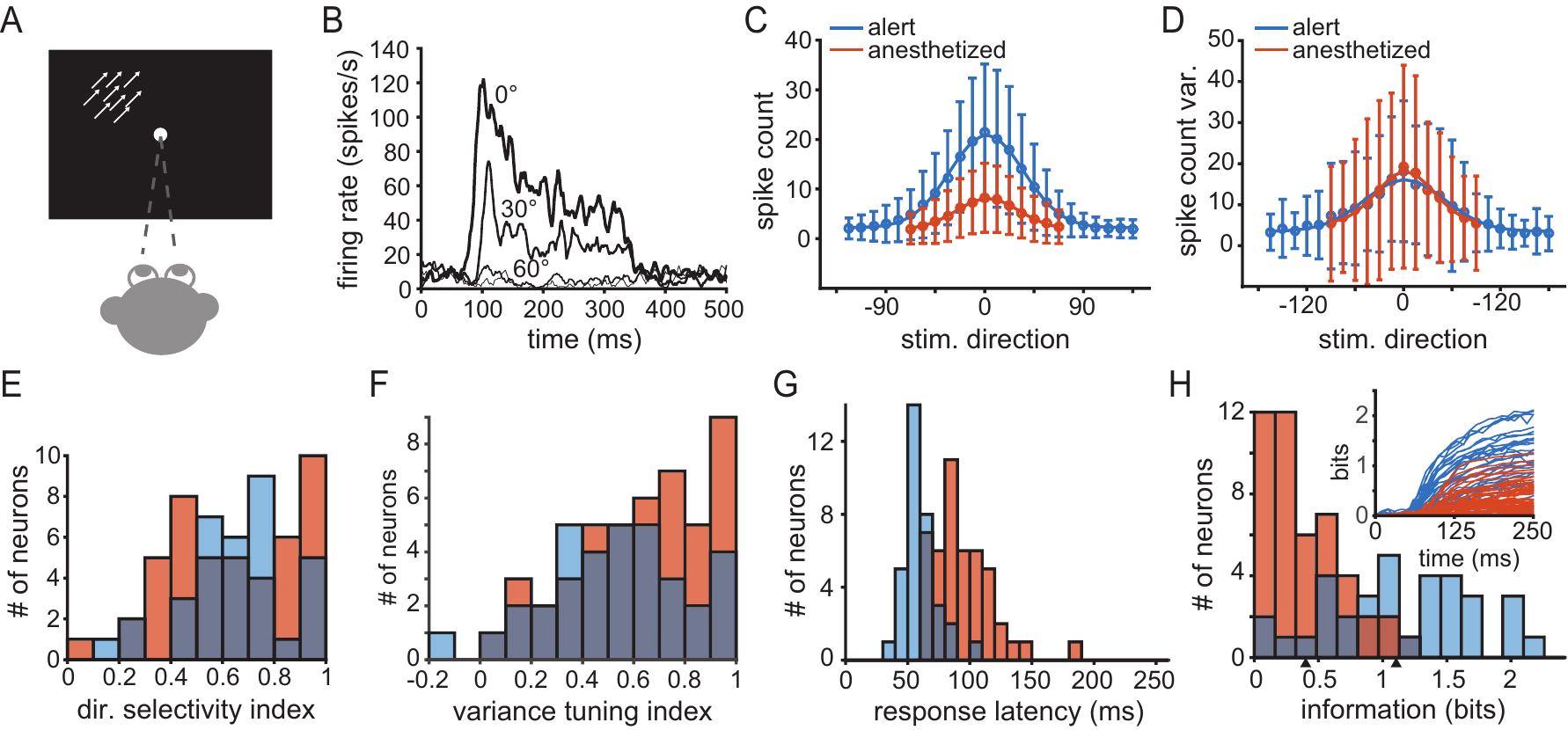}
	\caption{Response properties of MT neurons to constant motion stimuli. Blue indicates experiments on alert subjects and orange indicates experiments on anesthetized subjects throughout. (A) A cartoon of the experimental setup for measuring the response of MT neurons to steps of coherent random dot pattern motion. Monkeys maintained fixation while stimuli translated at constant speed in one of 13 or 24 directions behind a stationary aperture scaled to the excitatory receptive field. See Materials and Methods for detail. (B) Sample peri-stimulus time histogram (PSTH) from a representative unit showing the time course of firing rate for different motion directions, averaged over a 10 ms sliding window. Motion onset is at 0 ms and motion offset is at 250 ms. (C) Average spike count across neurons in the alert experiments (blue) and anesthetized (orange) by direction. Circles indicate mean, error bars indicate standard deviation, and the solid traces indicates a Gaussian best fit. Responses are aligned such that the preferred stimulus direction for each neuron is taken to be $0^{\circ}$. (D) Average spike count variance across populations of neurons in the same manner as (C). (E) Histogram of direction selectivity indexes, DI, across both populations. (F) Histogram of variance spike count tuning indexes. (G) Histogram of response latency distributions. (H) Distribution of single-unit mutual information (Shannon) values for alert (blue bars) and anesthetized (orange bars) states, based on spike count 250 ms after stimulus motion onset, Value are corrected for finite sample size (see Materials and Methods). Arrows below indicate average information for alert (1.11 bits) and anesthetized (0.40 bits). Inset: time course of the mutual information between the cumulative spike count and motion direction over time with respect to stimulus motion onset. Blue traces indicate alert-state units, orange traces anesthetized state.}
	\label{fig:stim_and_response}
\end{figure}

In order to test the impact of the network state on spiking statistics and stimulus encoding in cortical sensory neurons, we made extracellular recordings of isolated units from cortical area MT in monkeys. We compared responses in two states: alert and under light, opioid anesthesia (Fig \ref{fig:stim_and_response}A). Some of the data collected under anesthesia has been published elsewhere \cite{osborne_time_2004,osborne_2007,osborne_neural_2008}. In both states, MT neurons respond robustly to motion steps of random dot patterns, with firing rates that often peak at over 100 spikes/s for preferred directions (Fig \ref{fig:stim_and_response}B) and Gaussian-shaped direction tuning functions (Fig \ref{fig:stim_and_response}C). We presented motion steps in 13 (anesthetized condition) or 24 (alert condition) directions and repeated each stimulus $\sim 100$ times to estimate the distribution of spike counts (see Methods). In all experiments, the subjects maintained fixation during stimulus presentation. Under anesthesia, eye movements were suppressed with a paralytic agent. In the alert state, the subjects made small fixational eye movements that did not exceed a $2^{\circ}$ window around the fixation point. These movements are small compared to MT receptive field sizes and do not alter variability in motion-evoked responses in MT \cite{bair_1998}.

\subsection*{Anesthetic state modulates mean and Fano factor in MT neurons}
 MT neurons in alert, fixating primates show a pronounced increase in their stimulus-evoked mean firing rates compared to rates under light, opioid anesthesia (Fig \ref{fig:stim_and_response}C). We quantified the mean and variance of the response by counting spikes within a 250 ms window starting from stimulus motion onset on each trial. Although we focus on a single time scale, our results are consistent across a range of behaviorally relevant time scales for motion estimation (50-500 ms, see Supporting Information Fig S1). Mean counts were substantially higher in the alert condition than under light, opioid anesthesia which is consistent with comparisons between alert and anesthetized states in other systems (e.g.\ \cite{ecker_state_2014,white_2012,constantinople_bruno_11, chen_2011, alitto_2011, mcginley_2015, kisley_1999, aasebo_2017, gur_1997}). 
 
Like the mean count, the variance is tuned to the direction of motion in both states, with largely overlapping distributions of magnitudes (Fig \ref{fig:stim_and_response}D). On average, variance peaks at the preferred direction (rotated to $0^{\circ}$ in all figures) in both the alert (blue trace, Fig \ref{fig:stim_and_response}D) and anesthetized (orange trace, Fig \ref{fig:stim_and_response}D) populations and falls off with increasing angular separation from the preferred motion direction. Consistent with past studies, the change in state does not affect the direction tuning of mean rate. We computed a direction selectivity index, DI, (see Equation \ref{eq:direction_index}, Materials and Methods) and found that the population distribution of values were statistically indistinguishable (two-tailed t-test, $p=0.60$, Fig \ref{fig:stim_and_response}E). An analogous variance tuning index (see Equation \ref{eq:variance_index}) shows a similar tuning of the variance in both populations (two-tailed t-test,$p=0.09$, Fig \ref{fig:stim_and_response}F). Response latencies, estimated as the first point when the average response rose above baseline after motion onset and inspected manually for each neuron, decrease from $94 \pm 24$ ms (SD, n=46) under anesthesia (orange bars, Fig \ref{fig:stim_and_response}G) to $56 \pm 13$ ms (SD, n=34) in alert responses (blue bars). The latency difference is consistent with previously reported measurements \cite{kawano_1994,schmolesky_1998,raiguel_1999,raiguel_1989,aasebo_2017}. The increase in latency under anesthesia contributes to the reduction in estimated firing rate (Fig \ref{fig:stim_and_response}C), but firing rates are lower under anesthesia even when estimated in time windows aligned to response onset. 

Overall the impact of anesthesia is to lower signal (the mean rate) and maintain noise (count variance), suggesting that sensory information transmission may be impaired with respect to alert behavior, though no studies have directly measured this effect. To do so requires a population metric for the encoded information about motion direction. We address the possible population readout later in the Results, but first quantify changes in single neuron motion information. The mutual information between spike count and motion direction differs substantially between states (see Materials and Methods, Fig \ref{fig:stim_and_response}H). On average, MT units recorded in alert subjects encoded $1.11 \pm 0.57$ (SD, n=34, blue bars, Fig \ref{fig:stim_and_response}H) about direction compared to $0.40 \pm 0.32$ bits (SD, n=46, orange bars, Fig \ref{fig:stim_and_response}G) in anesthetized subjects, a statistically significant difference ($p=1.3*10^{-10}$, one-tailed t-test). The combined effect of the increase in response latency and reduction in information under anesthesia is that less information about motion direction is available over time. In Figure \ref{fig:stim_and_response}H inset we plot the time course of the mutual information between the cumulative spike count measured from motion onset and motion direction for each unit in both populations (see Materials and Methods). In both populations, stimulus information accumulates most rapidly with the first few spikes fired, but shorter latencies and higher overall firing rates in the alert state mean that more bits are available more quickly (blue versus orange traces, Fig \ref{fig:stim_and_response}H). Normalizing the mutual information by the response entropy reduced the difference between the two states but did not change the results. Changing the size of the time window over which counts are integrated does not recover the lost information under anesthesia and the difference in average coding capacities persists during stimulation. However, anesthesia does not entirely abolish the capacity of MT to encode information about motion direction. 

\begin{figure}[ht]
	\centering
		\includegraphics[width=1\textwidth]{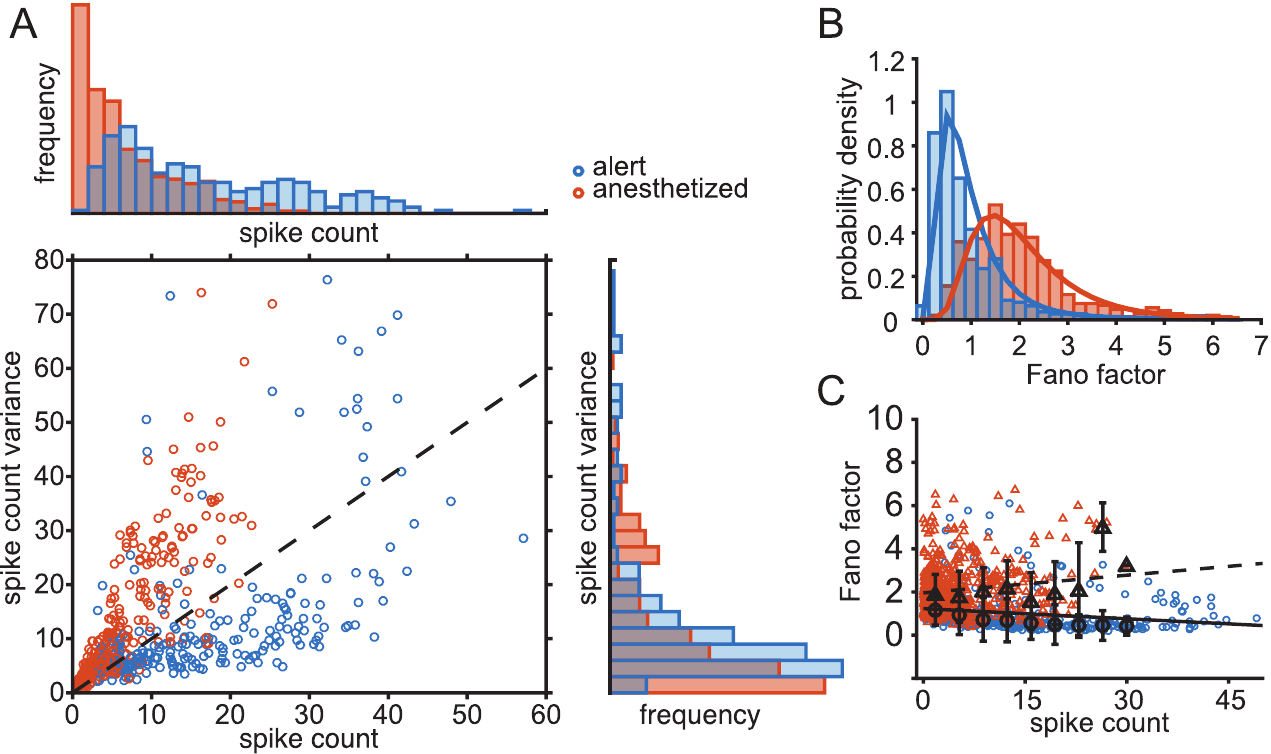}
	\caption{Fano factors differ significantly between MT responses recorded in alert versus anesthetized primates. (A) The relationship between spike count and spike count variance in alert (blue circles) and anesthetized (orange circles) experiments. Each circle indicates a single neuron under a single stimulus condition. Histogram above shows distribution of spike counts observed in both experiments. The histogram to the right shows the distribution spike count variance values in both experiments. Stimulus directions shown are within $45^\circ$ of the preferred motion direction, in $15^\circ$ increments. (B) Distribution of spike count Fano factors for alert (blue) and anesthetized (orange) experiments. Solid trace indicates the best-fit lognormal distribution for each case. (C) Distribution of Fano factors as a function of spike count for alert experiments (blue) and anesthetized experiments (orange). Black circles and triangles indicate the median Fano factor binned by spike count in alert and anesthetized experiments, respectively. Error bars show the standard deviation of Fano factors in each bin. The solid and dashed traces are the linear best fit for alert and anesthetized states, respectively ($r^2=0.05$ alert, $r^2=0.02$ anesthetized).}
	\label{fig:variance_to_mean}
\end{figure}

The maintenance of roughly equal spike count variance, but with higher mean firing rates in the alert condition and lower mean rates in the anesthetized condition, implies a substantial difference in the Fano factor (FF) between these two states. The state-dependence of the Fano factor is illustrated in Figure \ref{fig:variance_to_mean}. Very few neurons in either state display a Fano factor of 1 (dashed unity line in Fig \ref{fig:variance_to_mean}A). The Fano factors (FF) of most units measured in the alert state fall below 1 (blue symbols), while most units in the anesthetized state display FFs above 1 (orange symbols). The distribution of measured FFs over the anesthetized data shows a significant shift toward greater values compared to data from alert units, from a mean of $1.02 \pm 0.79$ (SD, n=34) to a mean of $1.82 \pm 0.84$ (SD, n=46) (Fig \ref{fig:variance_to_mean}B). These values are consistent with previous measurements and other areas, being mindful of the time window for counting spikes \cite{osborne_time_2004, churchland_stimulus_2010, kara_2000, gur_1997, gur_2006, ecker_state_2014, hires_2015}. Fano factors also show different dependencies on the count in the two states. Although values show a large degree of scatter, FF is flat or increasing with spike count in anesthetized data (orange and black triangles, dashed line, Fig \ref{fig:variance_to_mean}C). In the alert state, FFs tend to decrease slightly with increasing spike count (blue and black circles, solid line, Fig \ref{fig:variance_to_mean}C) such that neurons with higher firing rates tend to be more precise. Linear fits of rate to Fano factor (black traces, Fig \ref{fig:variance_to_mean}C) show these relationships, but low $r^2$ values (0.05 in alert, 0.02 in anesthetized) suggest that firing rate alone is not a great predictor of Fano factor. In neural data, Fano factors can depend on the time window over which spikes are counted \cite{teich_1996, averbeck_2003, osborne_time_2004, oram_2011}. We find that the time dependence differs between behavioral states.  Figure S1 plots the population mean FF values as function of window duration, with counting windows that expand from motion onset (main figure) or from response onset (inset). In the anesthetized state, the population mean FF increases with window duration (orange) but rises very slowly in the alert state (blue). Thus the separation between FF distributions plotted in Fig 2B increases for longer integration times. Overall, there is a marked shift in the scaling of variability with responsiveness in these two states. Alert state neurons have higher firing rates and higher rate precision relative to anesthetized responses, consistent with better direction discrimination during alert behavior. 

\subsection*{In alert subjects, Fano factor is tuned to motion direction}
All but one of the neurons showed directional tuning of the mean and 72 of 80 neurons showed directional tuning in the trial-to-trial variance (Fig \ref{fig:stim_and_response}C-F). However, the tuning of the variance does not simply follow the tuning of the mean, either in a one-to-one fashion, with a Fano factor of one, or with a constant factor not equal to one. In a linear regression between mean rate and variance, firing rate fails to explain much of the spike count variance measured in the alert experiments ($R^2=0.37$, see Materials and Methods for statistical tests used). A linear regression of spike count variance on firing rate explained more variance ($R^2=0.66$) in the anesthetized state. 

\begin{figure}[ht]
	\centering
		\includegraphics[width=1\textwidth]{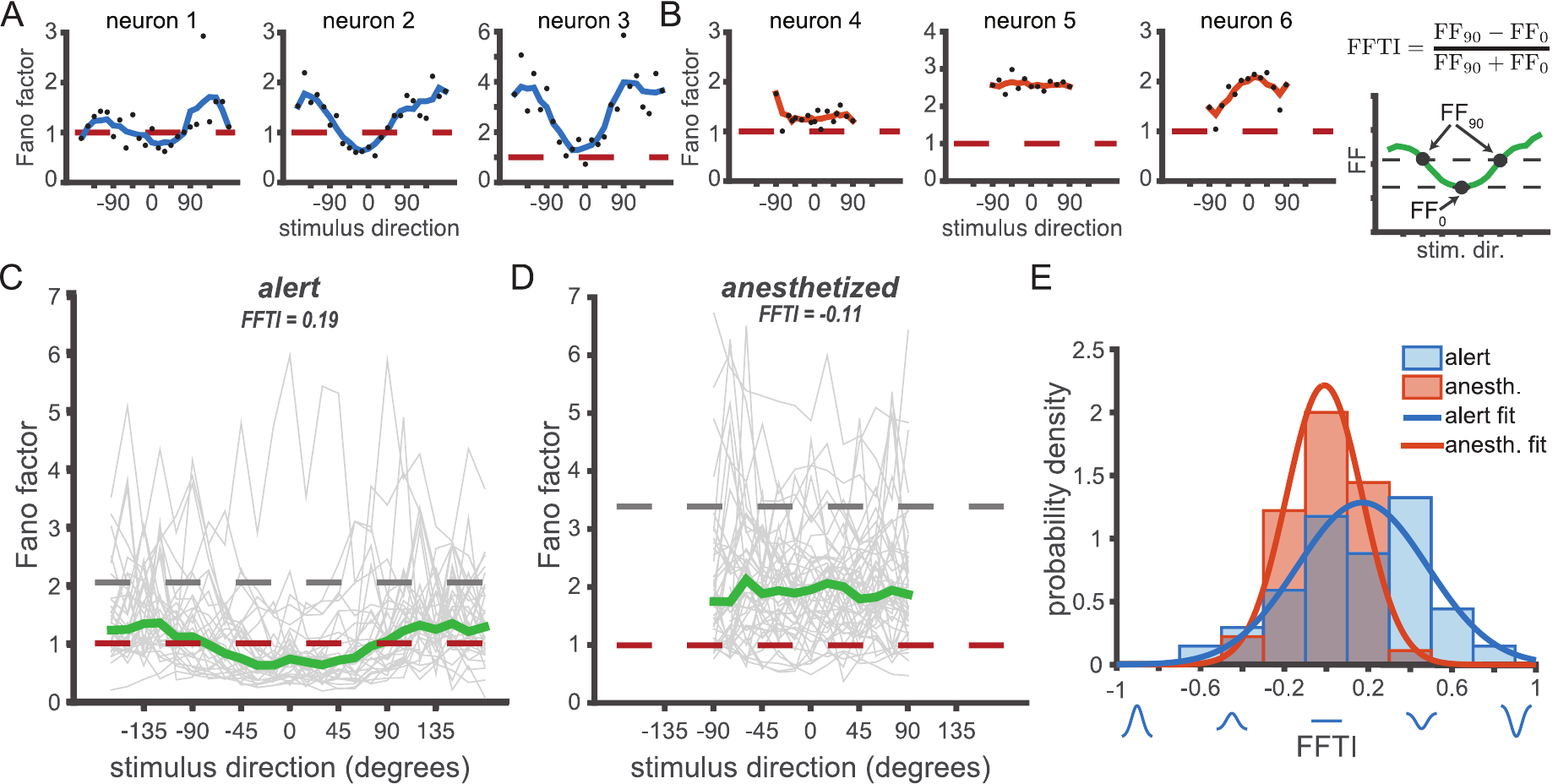}
	\caption{Fano factors show significant tuning to motion direction in the alert state. Fano factor across stimulus directions in example neurons from (A) alert and (B) anesthetized recordings in MT. Black dots indicate Fano factor by direction. Blue and orange traces show Fano factor values smoothed by a $30^\circ$ moving window for the alert and anesthetized states, respectively. Red dashed lines indicate $\mathrm{FF}=1$, as expected for Poisson firing. (C) and (D) Population FF tuning curves in each behavioral state. (C) Gray traces show the Fano factors versus stimulus direction for all neurons in the alert experiments. Stimulus directions are aligned such that the preferred stimulus direction of each neuron is $0^\circ$. The median Fano factor across the population (green) is significantly tuned for the preferred direction. The gray dashed line indicates the median Fano factor in response to a stationary, or `null', stimulus. Red dashed line is at $\mathrm{FF}=1$ as predicted for a Poisson process. (D) Same as (C) but for the anesthetized data. The median Fano factor is not tuned for stimulus direction. (E) Histogram distributions of Fano factor tuning indices (FFTI) from the alert (blue) and anesthetized (orange) experiments. Orange and blue traces are Gaussian best fit. The cartoon at top right shows how FFTI is calculated using the median trace from (C).}
	\label{fig:Fano_factor_tuning}
\end{figure}

Figures \ref{fig:Fano_factor_tuning}A and B show the qualitative difference in Fano factor tuning between three example units in the alert and anesthetized states, respectively. Although there are a range of FF tuning profiles measured within each state, the increased sharpness of FF tuning in the alert state is apparent. For each neuron recorded, we defined $\FFpref$, the Fano factor for the preferred stimulus direction of the neuron, and $\FForth$, the Fano factor for the orthogonal stimulus directions (rightmost panel, Fig \ref{fig:Fano_factor_tuning}B). We used these terms to define a Fano factor tuning index (FFTI) that captures the degree to which the FF depends on motion direction. 

\begin{equation}
\mathrm{FFTI}=\frac{\FForth-\FFpref}{\FForth+\FFpref}
\label{eq:FFTI_def}
\end{equation}

Equation \ref{eq:FFTI_def} is analogous to standard methods for quantifying direction selectivity in first-order response statistics like the rate \cite{priebe_tuning_2006,niebergall_2011}, and takes values from -1 to 1. Positive FFTI values indicate a decrease in Fano factor for the preferred stimulus direction relative to the off-preferred stimuli (``U-shaped'' tuning); negative FFTI values indicate an increase in Fano factor for the preferred stimulus direction (Gaussian-like tuning). The three alert-state example units in Figure \ref{fig:Fano_factor_tuning}A display ``U shaped'' tuning functions (FFTI values $> 0$) whereas the anesthetized examples in Figure \ref{fig:Fano_factor_tuning}B either lack direction tuning (FFTI near $0$, left and center panels) or show a more Gaussian-like profile (rightmost panel). We plot FF tuning functions for all units in Figure \ref{fig:Fano_factor_tuning}C (alert) and D (anesthetized). The gray lines connect the FF values measured at each direction for individual neurons. Green lines indicate the population median FF values, and the dashed gray line is the population mean FF computed for a stationary `null' stimulus, as a reference. Although there is a noticeable diversity in FF tuning profiles for individual units, particularly in the anesthetized state, the population medians (green lines, Figs \ref{fig:Fano_factor_tuning}C-D) reflect a clear state-dependent shift in tuning. Fano factors in the alert state are clearly tuned, whereas in the anesthetized state, the median FF is flat across directions. These results hold for mean FF (Fig S2) as well as median FF and for a variety of integration windows, including aligning to response latency rather than stimulus onset (Figs S3 and S4).

The distribution of FFTI values in the alert data is broader and more positive than in the anesthetized state (Fig \ref{fig:Fano_factor_tuning}E). Alert state MT units showed, on average, significantly greater FF tuning ($\langle\mathrm{FFTI}\rangle=0.172, \textrm{where } \langle \cdot \rangle$ indicates a mean over all units) compared to the anesthetized state ($\langle \mathrm{FFTI}\rangle=-0.012$) (two-tailed t-test, $p=0.0016$). We chose $\mathrm{FFTI}\geq0.2$ as a cutoff for a positively tuned Fano factor. In the alert state, 50\% of the units had positively tuned Fano factors (17/34). In the anesthetized state, only 13\% of the units had positively tuned Fano factors (6/45). One unit from the anesthetized group was excluded from the FFTI calculations as it did not emit any spikes to the orthogonal stimulus direction, and thus $\FForth$ was not defined. 

Overall, we observe a dependence of the Fano factor of the spike count on stimulus direction in the alert state. The tuning is U-shaped, with a dip at the preferred direction, and with the occasional presence of side-peaks at near-orthogonal directions (see e.g.\ Fig \ref{fig:Fano_factor_tuning}A-B). The Fano factor tuning we observe in the alert state using high Fourier bandwidth dot pattern motion is similar in shape and amplitude to that observed by \cite{ponce-alvarez_stimulus-dependent_2013} using low Fourier bandwidth drifting sine-wave grating stimuli. Our data from anesthetized subjects showed no significant stimulus dependence of the Fano factor (repeated measures ANOVA, $p=0.38$).

\subsubsection*{Firing rate changes do not explain differences in Fano factor tuning}

\begin{figure}[ht]
	\centering
		\includegraphics[width=1\textwidth]{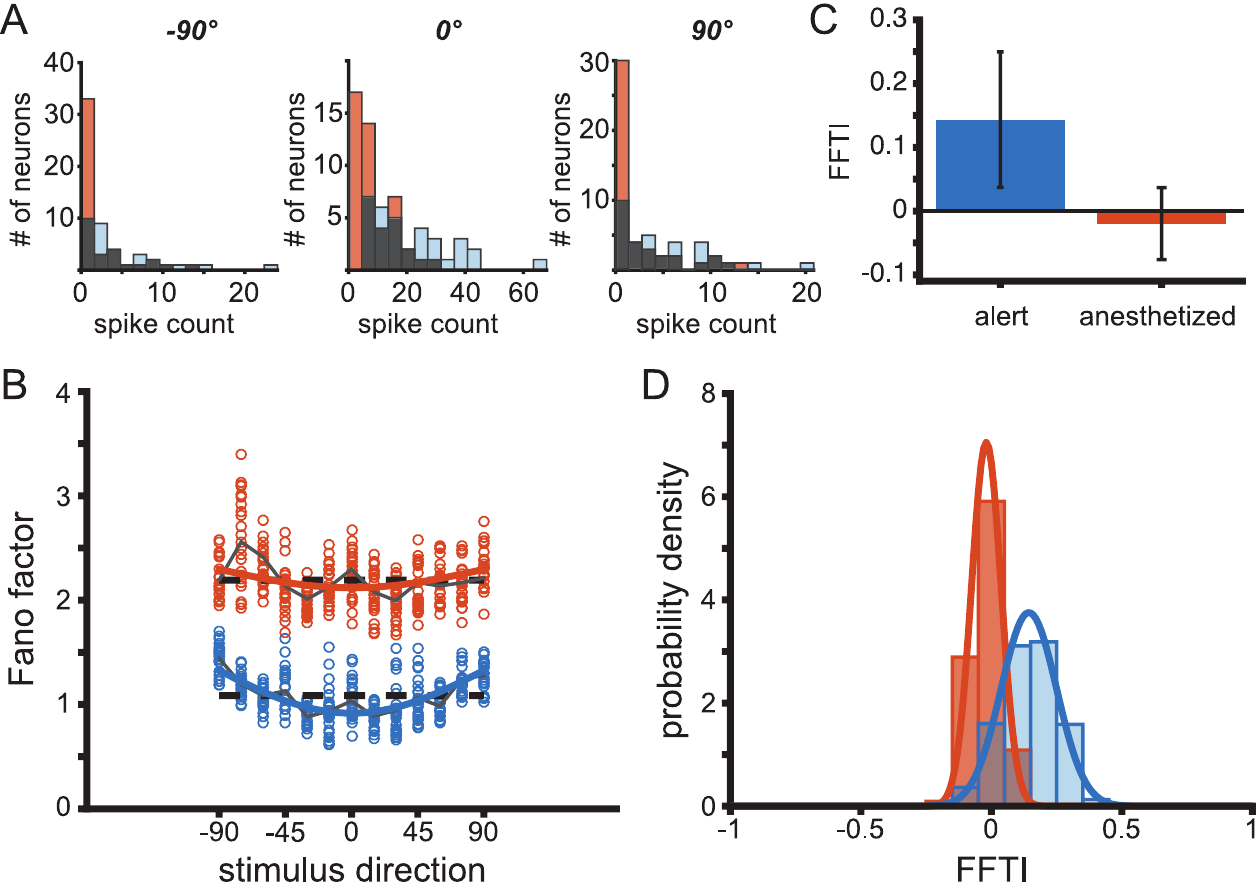}
	\caption{Mean-matched Fano factors preserve variability tuning. (A) Histograms of spike counts by neuron for a given stimulus direction (left: $-90^\circ$; center: preferred direction; right: $+90^\circ$) for alert (blue, throughout) and anesthetized (orange, throughout) recordings. The overlapping area is shown in gray. (B) Sample mean-matched Fano factors, with spike count distributions corresponding to the gray histograms in (A). Each circle is a sample mean. Dashed traces indicate the mean Fano factor across directions. Solid traces indicate the mean Fano factor across bootstrapped samples. (C) Fano factor tuning index (FFTI) of bootstrapped resamples for alert ($\mathrm{FFTI}=0.143$, $s.d.=0.106$) and anesthetized states ($\mathrm{FFTI}=-0.020$, $s.d.=0.057$). (D) Distributions of FFTI for mean-matched data samples. Solid traces indicate a best-fit Gaussian curve.}
	\label{fig:mean_matched}
\end{figure}

The firing rates in our alert MT recordings are roughly twice as large as those in the anesthetized recordings (Fig \ref{fig:variance_to_mean}A). Ideally, we would like to be able to exclude all effects changes of mean rate might have on observed Fano factors. One way to minimize the impact of rate differences is by analyzing subsets of the data samples that have matching firing rates, as in \cite{churchland_stimulus_2010}. The mean-matching method excludes data from the groups being compared in order to match their spike count distributions. Mean-matching does not maintain cell identity, and thus can only describe population-wide rather than cell-specific tuning. We generated histograms of the spike counts for both states at the preferred stimulus direction and at the orthogonal directions. We then randomly excluded points from each dataset until the histograms of spike counts in the alert and anesthetized states matched at each stimulus direction (gray shaded bars, Fig \ref{fig:mean_matched}A). We calculated the mean Fano Factor of this reduced data set for both stimulus conditions and both states. We calculated the population FFTI from the mean Fano factor at the preferred and orthogonal directions. This process was repeated one million times to average out effects of the particular random subset sampled on each draw.

The mean-matched analysis reveals that both the population-average shape of the FF tuning as well as the shift to higher FF in the anesthetized state are maintained (Fig \ref{fig:mean_matched}B). In this figure, the mean-matched population Fano factors were fit to cosine tunings (Fig \ref{fig:mean_matched}B). The best cosine fits corresponded to $\mathrm{FFTI}_\mathrm{alert}=0.156$ and $\mathrm{FFTI}_\mathrm{anesth}=0.039$ for the mean-matched populations, with significant tuning of the FF in the alert data. The state-dependent differences in Fano factor tuning for individual units within the sub-sampled population were also maintained after mean matching, with $\mathrm{FFTI}_\mathrm{alert}=0.143$ ($\mathrm{sd}=0.107$) and $\mathrm{FFTI}_\mathrm{anesth}=-0.020$ ($\mathrm{sd}=0.057$) (Fig \ref{fig:mean_matched}C). The distribution of FF tuning indices across sub-sampled data were also similar to the full data set, and recapitulated the shift to more positive FFTI in the alert state (Fig \ref{fig:mean_matched}D). 

An alternative possibility is that the state-dependent difference in FF stimulus tuning arises from differences in the scale of the FF between the two brain states. Fano factors in the anesthestized recordings were roughly $80\%$ higher than in the alert experiments, which affects the normalization factor in the tuning index calculation. With this in mind, we can examine the raw change in Fano factor, where $\Delta\mathrm{FF}=\FForth-\FFpref$. In alert recordings we see $\Delta\mathrm{FF}=0.326$ ($\mathrm{sd}=0.24$) and for anesthetized, $\Delta\mathrm{FF}=-0.09$ ($\mathrm{sd}=0.26$). Thus, the higher spike counts recorded in the alert state are not the source of the difference in Fano factor tuning.

It is notable that the qualitative relationship between Fano factor and firing rate differs in the alert and anesthetized recordings. Fano factors increase slightly with increasing mean spike count in the anesthetized state, and decrease with spike count in the alert state, over the same range of firing rates (Fig \ref{fig:variance_to_mean}C). This suggests that changes in mean spike count alone cannot entirely explain the observed differences in measured Fano factors and their tuning. 

\subsubsection*{Effect of tuning bandwidth on Fano factor tuning}
While differences in mean spike count fail to explain the differences in observed Fano factor tuning between states, it is possible that differences in tuning bandwidths lead to the observed differences in Fano factor tuning. If mean tuning curves are narrower or steeper in the alert state, this could result in different Fano factor tuning curves when normalized by the same variance tuning curves. Conversely, if the mean tuning curves are the same in both states, Fano factor tuning could arise from relatively broader or flatter variance tuning curves. We used an ANOVA to compare the widths of mean spike count tuning curves and spike count variance tuning curves in both states to the FFTI. There was no effect of tuning curve width or variance tuning width on FFTI (as is also evident in Fig \ref{fig:stim_and_response}E-F), and no interaction with state, suggesting that FF tuning arises from systematic differences in the structure of the variance rather than differences in the first-order tuning properties. 

\subsubsection*{Temporal correlations change with behavioral state}
\begin{figure}[ht]
	\centering
		\includegraphics[width=0.5\textwidth]{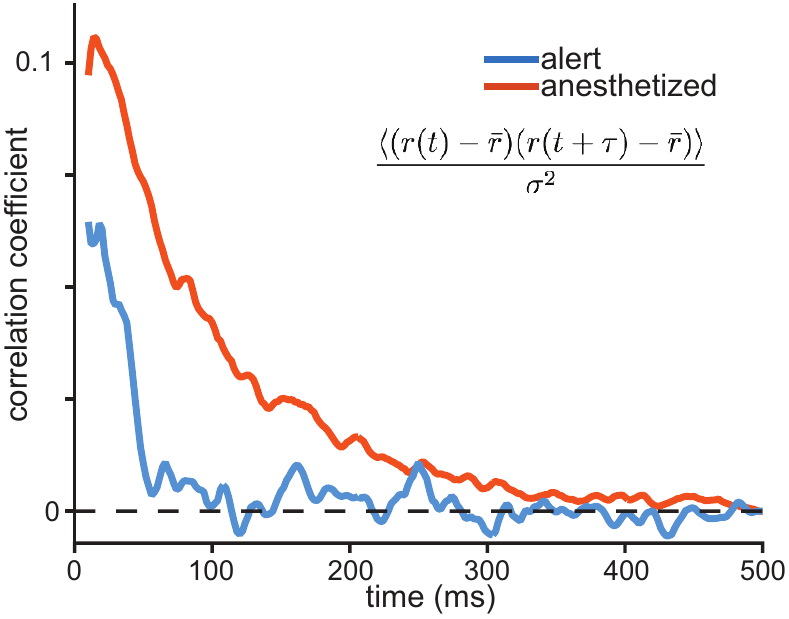}
		\caption{Behavioral state affects the spike-count autocorrelation function. Traces indicate the average temporal autocorrelation for spike count fluctuations in alert (blue) and anesthetized (orange) conditions. To calculate autocorrelation, spikes are binned in 2 ms windows and smoothed over a running average of 5 bins. Autocorrelation is normalized by variance and averaged over the population. The time course of autocorrelation is longer under anesthesia, decaying with an exponential time constant of 88 ms compared to 29 ms in the alert state.}
	\label{autocorr_fig}
\end{figure}

The temporal frequency of fluctuations in spiking differed substantially between the alert and anesthetized states. Under sufentanil anesthesia, neural excitability fluctuated more slowly such that deviations from the mean tended to accumulate during a trial. The temporal autocorrelation in spike count fluctuations displayed a higher peak with an exponential decay with a characteristic time constant of approximately 100 ms (Fig \ref{autocorr_fig}, orange curve). In contrast, alert-state correlations were weaker overall and no significant correlations in spiking were observed beyond 50 ms (blue curve, Fig \ref{autocorr_fig}). This difference in timescales contributes to the state-dependent difference in Fano factors. Longer, stronger temporal correlations in excitability create a larger count variance over a 250 ms time window, and therefore a higher Fano factor that grows with the expansion of the counting window \cite{osborne_time_2004}. The shorter timescales of correlation in the alert state create less variable counts, lowering the variance and the FF and reducing the dependence on the window duration. To the observer, fluctuations in spike count are indistinguishable from fluctuations in response gain, so an alternative description of MT activity is that, on timescales longer than 50 ms, gain fluctuations are smaller in the alert state than in the anesthetized state. We explore a gain-based model of state dependent changes in network activity below. 

\subsection*{A simple model accounts for changes in Fano factor tuning}
We created a simple model that could account for the observed state-dependent changes in FF tuning without relying on state-dependent differences in firing rates. The model is inspired by observations of decorrelation and enhanced variability in cortical responses in the alert compared to the anesthetized state. We reason that increased decorrelation with alertness could lead to lower relative variance in the neural response, and, hence, lower FF's. Models that include Fano factor tuning have been developed to explain the stimulus-dependent effects of neuronal response variability \cite{ponce-alvarez_stimulus-dependent_2013,franke_structures_2016,zylberberg_direction-selective_2016}, while others have modeled variability as rate-dependent, or due to stimulus-independent gain fluctuations \cite{scott_2012,goris_partitioning_2014}. Our model combines contributions from rate-dependent variability as well as rate-independent gain-fluctuations to reproduce the diversity of Fano factor tunings observed in both the alert and anesthetized data. 

A simple and common model of the variance in observed spike counts in cortical neurons is given by a Poisson process, or a Poisson mixture model \cite{goris_partitioning_2014}. A Poisson process is characterized by a Fano factor of 1. Including a multiplicative gain to describe the underlying rate of the Poisson process results in a super-Poisson spike count distribution, with a Fano factor that increases with firing rate. Under anesthesia, we found Fano factors larger than 1 (Fig \ref{fig:variance_to_mean}). These observations are consistent with a Poisson mixture model, and agree with previous experiments \cite{shadlen_variable_1998,goris_partitioning_2014}. Our recordings in the alert state, however, show a sub-Poisson spike count, with a Fano factor that decreases with increased firing rate. The inverse rate-dependence of the Fano factor was previously documented by \cite{ecker_state_2014} in V1. Can a single model account for responses in more than one behavioral state? The model we use captures the difference in the rate-dependence of Fano factors between states as well as the difference in Fano factor tuning.

We modeled stimulus averaged MT responses with Gaussian tuning functions $f(\theta)$, where $\theta$ is the stimulus direction. On any given trial, the mean rate is scaled by a multiplicative gain, $g$, yielding the underlying rate for the neuron on that trial 
\begin{equation}
\mu = f(\theta)*g.
\label{eq:rate_def}
\end{equation} 
The gain $g$ is taken to be a gamma-distributed variable with mean of 1. The spike count for a given trial is sampled from a Gaussian distribution with a mean value $\mu$, and a variance $\mu^\alpha$, where $\alpha$ is an intrinsic property of the cell that determines how variance scales with firing rate. For a fixed value of $g$, a value of $\alpha$ less than 1 corresponds to a Fano factor less than 1, and a value of $\alpha$ greater than 1 corresponds to a Fano factor greater than 1.

If we allow $g$ to fluctuate, the variance of the spike count $x$ is given by 
\begin{equation}
\operatorname{var}(x|\theta)=f(\theta)^\alpha\langle{g^\alpha}\rangle+f(\theta)^2*\operatorname{var}(g). 
\label{eq:variance_decomp}
\end{equation}
Because $\langle{g}\rangle=1$ and the variance of $g$ is relatively small, we can approximate $\langle{g^\alpha}\rangle\approx1$. Thus, we can approximate the spike count variance 
\begin{equation}
\operatorname{var}(x|\theta)\approx f(\theta)^\alpha+f(\theta)^2*\operatorname{var}(g).
\label{eq:variance_decomp_approx}
\end{equation}
For small values of $\operatorname{var}(g)$, \textit{i.e.}, small gain fluctuations, the variance is dominated by the first term. This results in a U-shaped Fano factor tuning for $\alpha<1$. As the gain fluctuations increase, the second term dominates the spike count variance, resulting in flat or Gaussian-shaped Fano factor tuning for the same value of $\alpha$. This model suggests that a difference in the amplitude of gain fluctuations is sufficient to reverse or eliminate the observed Fano factor tuning.

The gain fluctuation parameter also explains the qualitative dependence of Fano factor on increasing spike count observed in the anesthetized and alert states. For small gain fluctuations $(\operatorname{var}(g)\ll1)$ and $\alpha<1$, the Fano factor for a given firing rate, $r$, is $r^{\alpha-1}$, which decreases as $r$ increases. For stronger gain fluctuations, the Fano factor scales linearly with $r$, increasing with increasing rate. 

\begin{figure}[ht]
	\centering
		\includegraphics[width=1\textwidth]{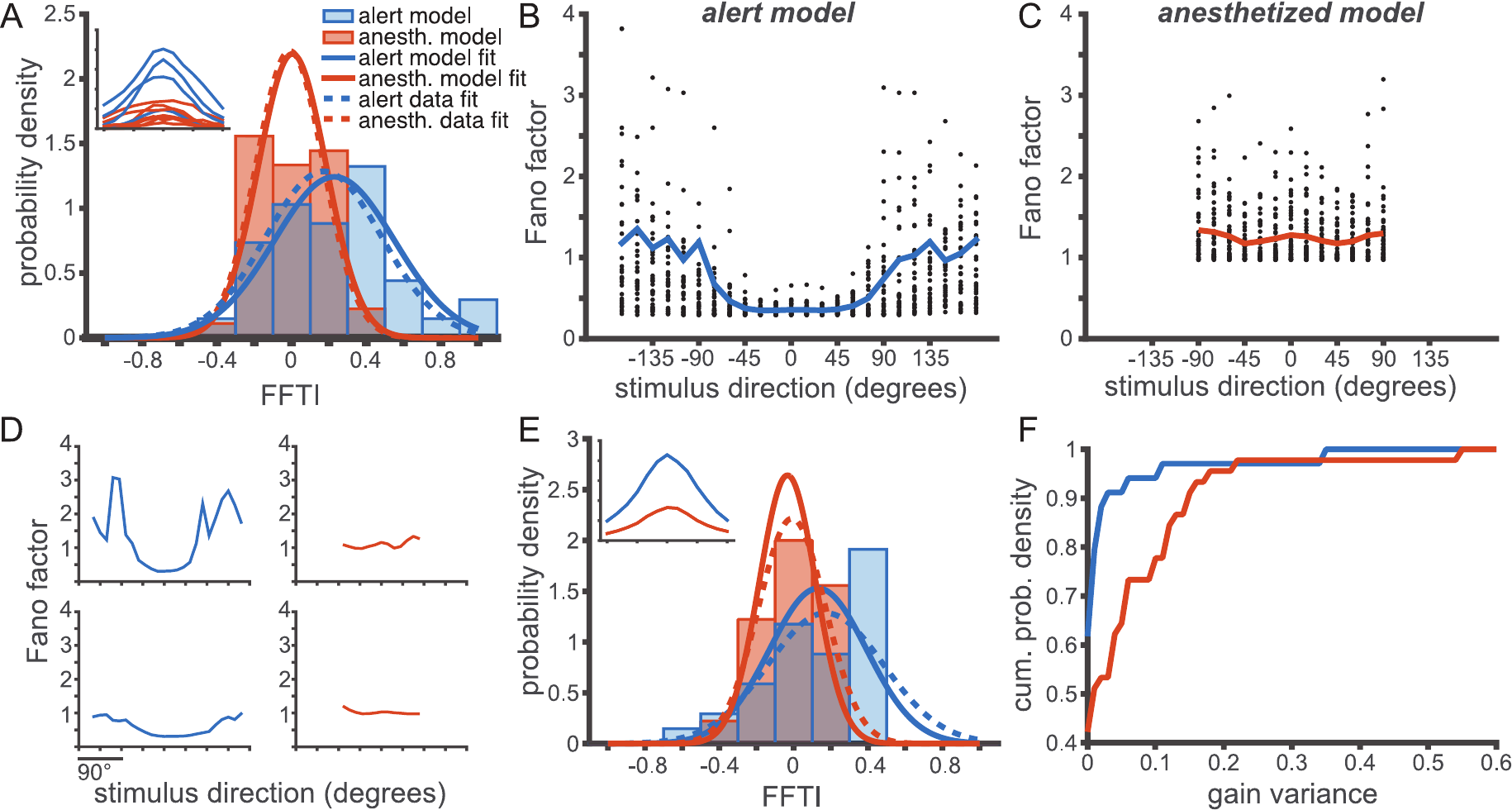}
		\caption{A single parameter, gain variance, can account for the observed changes in Fano factor tuning with behavioral state. (A-D) Model fitting captures differences in Fano factor tuning through changes in gain variance. The model is fit with the optimal $\alpha$ and $\operatorname{var}(g)$ for each population. (A) The distribution of FFTI for alert (blue, throughout) and anesthetized (orange, throughout) in the observed population (dashed trace) and the model values (solid trace). Inset: Sample tuning curves for the alert and anesthetized experiments. (B,C) The variance model predicts Fano factor direction tuning for the alert (B) and anesthetized (C) experiments. Compare model results to observed Fano factors in Fig \ref{fig:Fano_factor_tuning}C-D. Parameters are fit to match distribution of FFTI, but reasonably reproduce Fano factor tuning as well. (D) Sample Fano factor tunings generated by the best-fit models shown in (A-C). (E) Distribution of FFTI for alternate model fitting in which the spike count used for each neuron was replaced by the average tuning curve over all neurons recorded in each experimental condition. The $\alpha$ parameter is identical for all neurons and was chosen to minimize mean-squared error in the Kolmogorov-Smirnov distance between the observed and model FFTI distributions. The gain variance was fit separately for each neuron to match its observed FFTI (fit to minimize the mean-squared error). Inset: Mean tuning curves for the alert and anesthetized experiments. (F) The cumulative distribution of gain variance parameters from the model fit in (E), showing larger gain variance values in the anesthetized model.}
	\label{fig:ffti_model}	
\end{figure}

To fit this model to our data, we first use as input the set of best-fit Gaussian tuning curves, $f(\theta)_i$ where $i$ is the neuron label, measured in each behavioral state. Next, we fit an $\alpha$ and $\operatorname{var}(g)$ for each behavioral state to reproduce the observed FFTI distributions (Fig \ref{fig:ffti_model}A,B), where our error metric is the Kolmogorov-Smirnov distance between the observed and model distributions (see Fig S5). The distribution of Fano factor tuning indices generated by the model (Fig \ref{fig:ffti_model}A) were similar to those observed in the data (Fig \ref{fig:Fano_factor_tuning}D), indicating that the model can reproduce the observed data. The fitted values of the intrinsic variability parameter were $\alpha=0.31$ in the alert condition and $\alpha=0.74$ in the anesthetized condition. While these values are different, they both result in similarly U-shaped Fano factor tuning curves in the absence of large gain fluctuations. The model predicts a U-shaped Fano factor tuning for small gain fluctuations in both states, and inverted tuning for larger gain fluctuations. The gain variance parameters were $\operatorname{var}(g)=0.0094$ in the alert condition and $\operatorname{var}(g)=0.0732$ in the anesthetized condition, nearly an order of magnitude difference. It is this difference in gain variance that causes the qualitative difference in the Fano factor tuning between the two conditions. 

We also fit the model for the population-averaged mean-response tuning in each state, to test whether the differences in FFTI distributions could be wholly accounted for by changes in $\operatorname{var}(g)$, and not by changes in the distributions of mean tuning, $f(\theta)$. The average alert and anesthetized tuning curves are show in Fig \ref{fig:ffti_model}E, inset. The model parameters were then fit, such that all neurons in the same state had the same value of $\alpha$. The gain variance parameter $\operatorname{var}(g)$ was fit separately for each neuron to reproduce the observed distribution of FFTIs. This model again predicts a distribution of gain fluctuations that is much larger in the anesthetized condition than in the alert condition, and is centered on a significantly larger mean gain variance value (Fig \ref{fig:ffti_model}F). In both models, whether the tuning curves are heterogeneous or identical, the proportion of neurons with high values of gain variance is much greater in the anesthetized condition. 

To confirm that the relative shift in the FFTI distributions was not dominated by differences in $\alpha$ in each behavioral state, we fit the same model using an best-compromise value of $\alpha=0.53$ for every neuron in both behavioral states. This $\alpha$ was chosen so as to minimize the squared sum of the KS distance between the model and observed FFTI distributions in each state. The results were comparable to those shown, and the difference in gain variance between states was yet larger (see Fig S6). The complementary model, in which a single gain variance parameter is used for both behavioral states and the $\alpha$ parameter is left to reproduce a shift in the FFTI distribution fails to capture even the qualitative properties of the FFTI distributions (see Fig S7). The anesthetized data show significant FF tuning, larger than in the alert state, in this model. This indicates that the observed FFTI distributions might be primarily modulated by changes in a single parameter, the variance in the gain distribution.

Finally, we fit a model in which each cell's Fano factor tuning curves were fit using individual $\alpha$ and gain variance parameters (see Fig S8). The distribution of $\alpha$'s was broad, with no significant difference between behavioral states (see Fig S9). The gain variance parameters, however, were again an order of magnitude higher for the anesthetized data (see Fig S10).

\subsection*{Impact on information transmission}
It seems reasonable to assume that a higher Fano factor in the anesthetized state will result in lower rates of information transmission and stimulus discriminability in these neural populations. Conversely, the reduction in relative variability at the preferred direction for most cells recorded from alert, behaving subjects seems to imply enhanced information transmission and decoding. To test this intuition, we simulated responses of MT neuron populations and varied the stimulus-dependence of the noise via our model. Our goal is to compare, qualitatively, the effects of Fano factor tuning on stimulus discriminability. We do not aim to perform an exhaustive exploration of the potentially related effects of variance tuning on correlation and discriminability. Instead we ask how discriminability is affected by FF tuning, all other aspects of the population code being equal. We quantify population encoding performance via the Fisher information (FI) metric, which determines the bound on the performance of an unbiased estimator reading out motion direction from the population code. 

The neurons in our data were, for the most part, recorded independently, so we must proceed with caution in evaluating the absolute magnitude of stimulus discriminability in our population models. Information-limiting correlations are always present in a real neural population, and set hard bounds on the stimulus discriminability, see e.g.\ \cite{ecker_2016}. The magnitude of these correlations can be so small as to be experimentally undetectable, yet they can have an outsized impact on the population Fisher information \cite{moreno_2014,kanitscheider_2015,pitkow_2015,zylberberg_2017,ecker_2016}. We can, however, infer the maximal size of these information-limiting correlations by using behavioral data on thresholds for motion discrimination in primates \cite{osborne_2005, osborne_2007, mukherjee_2015}. If the animal can distinguish motion direction down to a carefully measured threshold, the information in the neural population that drives this behavior must have at least that much discriminability. We leverage the nearly noiseless information transfer between motion estimation and pursuit eye tracking to estimate the information that the brain itself recovers from the MT population \cite{osborne_2005, osborne_2007, medina_lisberger_2007, mukherjee_2015, lee_etal_2016}. We bound the information capacity in our model MT populations at this behavioral benchmark, which sets a strong upper bound on the effects of information-limiting correlations. 

\begin{figure}[ht]
	\centering
		\includegraphics[width=1\textwidth]{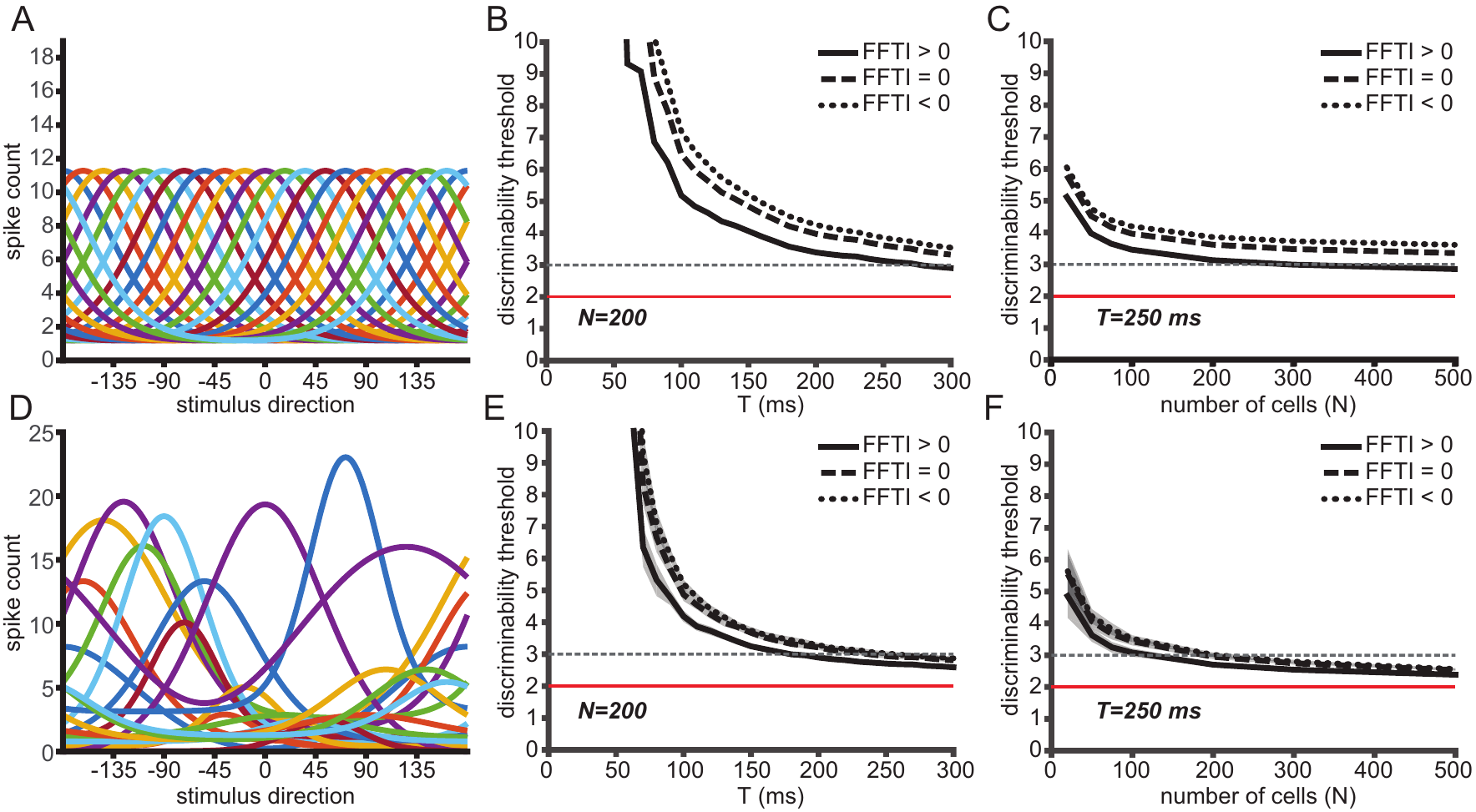}
		\caption{Fano factor tuning and heterogeneity both contribute to lower discriminability thresholds in MT populations. (A) Model population with 20 homogeneous tuning curves. (B) Cramer-Rao bound in homogeneous population models of 200 neurons over time with short-range correlations ($c_{max}=0.1$, $\langle c\rangle=0.0438$) and information-limiting correlations ($\epsilon=4$). Tuning curves are fit to the average cumulative response from motion onset up to time T. Black traces show performance of models with varying stimulus-dependent variance. The solid trace is $\textrm{FFTI}>0$, the dashed trace is $\textrm{FFTI}=0$, and the dotted trace is $\textrm{FFTI}<0$. The dashed gray line indicates the stimulus discriminability threshold for smooth pursuit behavior in macaques 125 ms after pursuit initiation. The red trace indicates the bound on discriminability at $2^\circ$ set by the information-limiting correlations. (C) Same as in (B) but for neuronal populations of different sizes with first-order statistics matched to the average response at time 250 ms after motion onset. (D) Sample population of 20 heterogeneous tuning curves drawn from measured tuning curves in recorded neurons. (E,F) Same as in (B,C) but for heterogeneous populations. The shaded areas show the standard deviation of the Cramer-Rao bound.}
	\label{fig:population_model}
\end{figure}

For completeness, we also impose stimulus-dependent pairwise correlations in our model population, by generating synthetically-correlated population responses based on published MT data (see Methods and \cite{bair_2001,huang_2009}). In our simulations, pairwise correlation levels, $c$, peaked at 0.1 and fell off with the angular distance between preferred directions, $d$, according to a von Mises function 
\begin{equation}
c(d) = c_\textrm{max} \frac{e^{\kappa(\cos{(d)}+1)}-1}{e^{2\kappa}-1}
\label{corr_equation}
\end{equation}
with width parameter, $\kappa=1$, corresponding to a half-width at half-max $= 64^\circ$. For the results shown in Figure \ref{fig:population_model}, the maximum pairwise correlation $c_{\textrm{max}}$ was chosen to be 0.1, with an average pairwise correlation of $\langle c\rangle=0.0438$. In the analysis we present here, we compare identically coupled populations in order to isolate how changes in the FF tuning affect discriminability. We note that changes in FF tuning may go hand-in-hand with changes in the pairwise correlation structure in the population. These concomitant changes could lead to qualitatively different behavior than what we observe with fixed correlations. 

Information-limiting correlations take the form \cite{moreno_2014}
\begin{equation}
 \mathbf{\Sigma_\epsilon}(\theta)=\mathbf{\Sigma_0}(\theta)+\epsilon\mathbf{f}'(\theta)^T\mathbf{f}'(\theta),
\label{differential_corr}
\end{equation}
 where $\mathbf{\Sigma_0}$ is the initial covariance matrix, $ \mathbf{f}'(\theta)$ is the derivative of the tuning curves, $\epsilon$ is a constant, and $\mathbf{\Sigma_\epsilon}$ is the covariance with information limiting correlations. The linear Fisher information for the population is calculated as in equation \ref{eq:fisher_information}. If the Fisher information associated with the covariance $\mathbf{\Sigma_0}$ is $J_0$, the Fisher information associated with $\mathbf{\Sigma_\epsilon}$ is
\begin{equation}
 J_\epsilon =\frac{J_0}{1+\epsilon J_0}.
\label{limited_fisher_info}
\end{equation}

We quantify performance by the square root of the Cramer-Rao bound, given by
\begin{equation}
 \frac{1}{\sqrt{J_\epsilon}}=\sqrt{\frac{1}{J_0}+\epsilon},
\label{cramer_rao}
\end{equation}
which represents the lower bound on the standard deviation of an unbiased estimator. As $J_0$ increases with population size, the decoder performance is bounded by $\sqrt{\epsilon}$. 

The MT population response is the substrate for motion perception and for motion-driven behaviors like smooth pursuit eye movements (reviewed in \cite{Lisberger_2015}). Pursuit is a tracking behavior that rotates the eye along with a moving target in order to stabilize its retinal image. The precision of pursuit and perceptual thresholds for discriminating motion direction are well-matched, with values of about $3^\circ$ after 125ms and within the ``open-loop'' period before feedback effects take hold \cite{groh_1997, born_2000, osborne_time_2004, osborne_2005,osborne_2007, stephens_2011, hohl_2013, mukherjee_2015}. The precision of the behavior suggests that little noise is added downstream of the visual estimate decoded from the MT population. With longer viewing periods or after longer bouts of pursuit, behavioral discrimination can reach about $2^\circ$ precision \cite{mukherjee_2015}. We chose $\epsilon = 4$, corresponding to a precision of $2^\circ$, as a benchmark to evaluate model performance in Figure \ref{fig:population_model}, setting this as the floor for the discriminability in the neural population. We then evaluated when otherwise identical models with and without FF tuning reached the open-loop behavioral threshold of $3^\circ$. 

We first measure the Fisher information in a homogeneous population with identically shaped tuning curves (Fig \ref{fig:population_model}A). The first-order response properties of the simulated populations were matched to the population-averaged statistics of MT neuron responses measured in alert macaques in the first 150 ms following the onset of stimulus motion. We varied this averaging window systematically to determine how it affects the resulting decoding performance of the model. The mean tuning curve was fit to a von Mises function as in \cite{ecker_state_2014,zylberberg_direction-selective_2016} and rotated such that the preferred directions evenly tiled all directions.

In order to separate the effects of the magnitude of the spike-count variability from the stimulus-dependent part of the variability, we imposed various Fano factors tunings on the populations while keeping the average Fano factor across all directions constant (mean FF=1). The Fano factor tunings were modeled as von Mises functions and imposed directly, rather than using the spike count model described above, so that the average Fano factor could be held constant. The positive (U-shaped) and negative (inverted-U) tuning of the stimulus-dependence of the Fano factor are symmetric by reflections over the line FF=1.

In the homogeneous population, the population with the positive FFTI performed better than a population with either a flat Fano factor tuning or a negative FFTI (Fig \ref{fig:population_model}B,C). 
We looked at the time course of decoding performance by finding the average tuning curves after cumulative intervals of time after motion onset (Fig \ref{fig:population_model}B). We assumed a population size 200 neurons and $c_{max}=0.1$. We estimated the time after motion onset when the square root of the Cramer-Rao bound for these populations would cross the $3^\circ$ threshold. We asked whether our model populations could reach that direction precision within the behaviorally relevant timescale. The population with $\textrm{FFTI}>0$ crossed the threshold after 277 ms, while the $\textrm{FFTI}=0$ and $\textrm{FFTI}<0$ populations did not reach the $3^\circ$ threshold within the analysis window. 

We also looked at the population size required to reach the behavioral discrimination criterion within a 250 ms analysis window. The homogeneously tuned model population with $\textrm{FFTI}>0$ crossed this threshold with 294 neurons (Fig \ref{fig:population_model}C). The populations with $\textrm{FFTI}=0$ and $\textrm{FFTI}<0$ in which the variability was structured to give a constant Fano factor ($\textrm{FFTI}=0$) did not reach the $3^\circ$ discriminability threshold with a population of 5000 neurons, only reaching $3.12^\circ$ for $\textrm{FFTI}=0$ and $3.42^\circ$ for $\textrm{FFTI}<0$. Positive FF tuning, meanwhile, resulted in significantly higher discriminability at this large $N$, reaching $2.48^\circ$ for 5000 neurons. 

Real cortical populations are heterogeneous, and that heterogeneity is expected to improve overall coding \cite{shamir_implications_2006,osborne_neural_2008}. We introduced heterogeneity into our simulations by sampling with replacement from the measured tuning curves in each behavioral state and randomly reassigning preferred directions uniformly (Fig \ref{fig:population_model}D-F). The tuning curves were again fit to von Mises functions that matched the cumulative spike counts assessed at a range of intervals following stimulus motion onset and the preferred directions were spaced evenly to tile the space of stimulus directions. We found that adding this heterogeneity did improve coding performance, as assessed by the Cramer-Rao bound. Fewer neurons were required to reach behavioral performance, and this bound could be reached more quickly after motion onset in the heterogeneous populations (Fig \ref{fig:population_model}E-F). Short-range correlations and stimulus-dependent variability were imposed in the same manner as with the homogeneous population.

We found that the effect of stimulus-dependent variability was smaller in the heterogeneous population model as compared to the homogeneous one, but the same qualitative trend was observed. The heterogeneous population sampled at 250ms after motion onset and with $\textrm{FFTI}>0$ required only 126 neurons to reach a Cramer-Rao bound of $3^\circ$, while the population with constant Fano Factor required 194 neurons (Fig \ref{fig:population_model}F). The population with $\textrm{FFTI}<0$ required 219 neurons to reach the same level of stimulus discriminability (Fig \ref{fig:population_model}F). 

Estimating the time course of stimulus discriminability in populations of 200 heterogeneous neurons, the $\textrm{FFTI}>0$ population reached the $3^\circ$ discriminability threshold in 173 ms while the $\textrm{FFTI}=0$ population reached the same threshold in 237 ms and the $\textrm{FFTI}<0$ population reached the threshold in 257 ms (Fig \ref{fig:population_model}E). Once again, U-shaped tuning of the Fano factor, as observed experimentally, allows the threshold on behaviorally relevant direction discrimination levels to be reached with fewer neurons in a shorter amount of time after motion onset. 

\section*{Discussion}
The brain functions across a wide range of network states that encompass levels of arousal and attention, yet little is known about how behavioral state affects sensory discrimination. Much of our historical understanding of the nature of visual coding arises, of necessity, from experiments under anesthesia, mimicking a stage of sleep \cite{hubel_1962,maunsell_functional_1983,brown_2010}. Those observations continue to inform modern neuroscience because the response characteristics of cortical neurons, such as tuning curves, remain consistent across network states \cite{treue_1999,mcadams_1999,cook_maunsell_02,alitto_2011}. But there are clear state-dependent differences in sensitivity, background firing rates and pairwise correlation structure with alertness and attention \cite{treue_1999,cohen_2009,ecker_state_2014,white_2012}. Attention, for example, decorrelates local cortical populations \cite{cohen_2009} and increases firing rates for preferred stimuli \cite{treue_1999,martinez_2004}, enhancing signal-to-noise ratio in the cortical network and improving stimulus detection and discrimination \cite{shadlen_variable_1998,butts_2006}. The fine spatial \cite{mitchell_2009} and temporal \cite{ghose_2002,vanede_2011,cohen_2009,luck_1997} scales over which attention can operate suggests a localized control mechanism. Here we show that a simple model operating at the level of individual neurons can reproduce many features of the state-dependent variance-to-mean changes we observe in sensory cortex.

We have used a combination of physiological data analysis from cortical area MT alongside modeling to characterize the statistics of cortical responses under two candidate behavioral states: attentional alertness required for maintaining fixation during a visual experiment, and a quiescent state induced by the opioid anesthestic agent, sufentanil. A characteristic effect of opioid anesthesia is to increase cortical wave activity, low frequency spatiotemporally structured activity modulations (e.g. \cite{brown_2010,townsend_2015}). While the scale of wave activity is increased, it is not observed to be unnaturally structured. Similarities in the correlations between functionally connected brain areas in the quiet awake and lightly anesthestized states suggests that general activity patterns under anesthesia can mimic network states during active behavior \cite{mohajerani_2013,harris_2011,vincent_2007}.

The larger Fano factors we observe under anesthesia are not likely to be caused by cell-intrinsic mechanisms within area MT. Intrinsic noise in spike generation is quite small, even in cortex. For example, Mainen and Sejnowski (1995) \cite{mainen_sejnowski_1995} showed that when isolated from the network by current clamp, cortical neurons fire precisely, with coefficients of count variation (SD/mean) of 0.1 for a constant input current and of 0.05 for a dynamic stimulus. The apparent variability of cortical spiking likely arises from sources of variation not controlled by the experimenter such as random seeds in visual stimulus generation, slow-wave modulation of excitability, and amplification of small noise sources by recurrent \textit{but deterministic} networks, and idiosyncratic eye movements \cite{bair_koch_1996, vanvree_somp_1996, shadlen_newsome_1998, vogels_abbott_2005, gur_1997, gur_2006, churchland_stimulus_2010}. 

The strongest evidence for the origin of the shift in FF level between the anesthetized and alert states is the shift in the level and time span of correlations in excitability shown in Fig \ref{autocorr_fig}. The increase in temporal correlations in single unit spiking is consistent with the increase in population-level low frequency oscillations under sufentanil anesthesia. The shift from a flat FF to a direction tuned FF in the alert state suggests that count variation has a different origin in the two states. Direction-tuned variation could be inherited from V1 in a feed-forward manner by the same mechanism that creates direction tuning of the firing rate. Another possibility would be a top-down contribution from direction-tuned neurons in MST. The emergence of widespread oscillations during anesthesia swamps direction-tuned variability by inducing, in our model, large gain fluctuations.

Overall, these gain fluctuations lower the precision of MT responses and degrade stimulus discriminability compared to the alert state. Models of MT activity that can describe the shift in variance and mean firing rates between the alert and anesthetized states have an additional constraint. We find, as did \cite{white_2012,ponce-alvarez_stimulus-dependent_2013}, that response variance is stimulus-dependent and has its own tuning function that is similar, but not identical, to that of the mean count. A measure of response precision, the Fano factor, acquires stimulus tuning in the alert state and becomes stimulus-independent under anesthesia. Significant tuning of the Fano factor during alert behavior may enhance stimulus readout. Lower variance at and around the preferred direction of each neuron in a population leads, unsurprisingly, to a finer discrimination threshold, as estimated via the Fisher information with important constraints set by a behavioral estimate of the information-limiting correlations. Fewer cells are needed to achieve the same level of direction discriminability, and stimulus information is available earlier in populations that have this kind of Fano factor tuning. In all of these analyses, we modeled an identical correlation structure in the neural population. Of course, behavioral state may also affect correlations as well as FF tuning, so the results we present here should be viewed as a test case to explore how FFTI affects discriminability. In the real brain, competing effects may negate the relative benefits of positive FF tuning, and testing this requires recording from larger neural populations. 

Using behavioral thresholds to estimate the size of information-limiting correlations may be a useful strategy in modeling populations of neurons when large simultaneous recordings are not available or experimentally feasible. Even if simultaneous recording is possible, say for pairs of neurons, evaluating the size of information-limiting correlations in a larger distribution of non-information-limiting correlations can be difficult if not impossible \cite{ecker_2016}. When behavioral estimates are available, they may allow for the dissection of these different types of correlation. Behavioral discrimination thresholds may vary with brain state and should be separately estimated in each state to disentangle the effects of a shift in information-limiting correlations and changes in FF tuning. 

The fact that the response variance is not tied to the mean firing rate violates the usual assumption that cortical spiking has Poisson statistics. A Poisson process yokes mean and variance together to maintain a Fano factor of 1, or values slightly less than 1 when refractoriness is revealed at high firing rates \cite{mitchell_2007}. Neural deviations from idealized Poisson behavior are well documented \cite{bair_2001,buracas_1998,stevens_1998,kara_2000,maimon_2009} and these non-Poisson effects are known to be important for accurate modeling of neural response \cite{keat_2001,pillow_2003}. Deviations from Poisson behavior are particularly acute in our data and represent a strong constraint on a feasible model of cortical activity that can generalize to different behavioral states.

We show that a model that incorporates a state-dependent shift in gain variance alone can reproduce the changes in the variance level and variance tuning observed in MT. Other recordings and models of attentional effects on neuronal firing, particuarly in area MT, focus on tuning-dependent shifts in the gain of mean responses \cite{reynolds_2009,ni_2012,ni_2017}. Here, we focus on the effects of variance in the gain on the stimulus tuning of the Fano factor. Using the cortical gain model proposed by \cite{goris_partitioning_2014}, we are able to show that a change in the gain variance model can not only explain overall shifts in the Fano factor, but also reproduces tuning of the Fano factor in the alert state, when gain variance is low. This suggests that a higher gain variance state underlies the observed flat tuning of the FF under anesthesia. This aligns with results that suggest that anesthesia corresponds to a more synchronous mode of brain coupling \cite{brown_2010,castelo_1998,chen_2011}. These results point to a simple physiological mechanism that achieves the shift in response statistics with alertness. A recent study in rat V1 shows effects of anesthetic state on Fano factors and FF tuning to the stimulus period \cite{white_2012}, adding general support to our observation that noise is suppressed in the alert state. 

The active suppression of gain fluctuations during wakefulness may be directed specifically at those neurons encoding stimulus variables in an active task (here, fixation). This suggests a simple knob that attentional modulation can turn to drive more reliable decoding of the stimulus. Reducing gain fluctuations may require processes, such as activation of inhibitory networks, that are metabolically costly and, thus, are only engaged when needed to maximize sensory discrimination during active behavior. 

\section*{Materials and Methods}
\subsection*{Ethics statement}
All procedures were performed in compliance with Institutional Animal Care and Use Committee guidelines at the University of California, San Francisco, and the University of Chicago. All procedures complied with guidelines for animal welfare in accordance with the recommendations of the Weatherall report, ``The use of non-human primates in research''. During anesthetized experiments, anesthesia was maintained with continuous infusion of the opioid sufentanil, paralysis was induced with vecuronium bromide to minimize eye movements, and midazolam was administered periodically. Pain responses were monitored at 15 minute intervals and additional analgesics were used if necessary. Similar alert motion-step experiments were performed with two adult male monkeys (Macaca mulatta) that maintained fixation during visual stimulus presentation. Animals were pair-housed when possible and had daily access to enrichment activities and play areas. Pre-study instrumentation with a head stabilization post, an eye coil, and a recording chamber was performed under isoflurane anesthesia using sterile surgical technique and post-operative analgesia with buprenorphine. Animals were trained over a period of time to acclimate them to the laboratory environment. Daily experiments involved seating the monkey in a plastic ``chair'' in front of a visual display in a dimly lit room. The task required fixation of a spot target within 2 degrees throughout the 2-3 second trial to obtain a juice reward.

\subsection*{Experimental methods}
We made extracellular single-unit microelectrode recordings of MT neurons in both alert, behaving and anesthetized monkeys. All procedures were performed in compliance with \textit{Institutional Animal Care and Use Committee} guidelines. The ``anesthetized motion-step experiments'' consisted of recordings from four adult male macaques (\textit{Macaca fasicularis}). Animals were implanted with a head-restraint and a craniotomy was performed under isoflurane anesthesia using sterile technique. During the anesthetized experiments, anesthesia was maintained with continuous infusion of the opioid sufentanil, and paralysis was induced with vecuronium bromide to minimize eye movements, and midazolam was administered periodically. Pain responses were monitored at 15 minute intervals and additional analgesics were used if necessary. Unit recordings were made using tungsten-in-glass microelectrodes. Visual stimuli comprised randomly drawn patterns of white dots that moved coherently within a stationary aperture against the dark screen of analog oscilloscopes (models 1304A and 1321B, P4 Phosphor; Hewlett-Packard, Palo Alto, CA). The size and position of the stimulus aperture was chosen to maximally excite each isolated unit, as was motion speed. The direction of stimulus motion was pseudo-randomly chosen from a set of at least 13 directions that spanned $\pm90^{\circ}$ around the preferred direction including $15^{\circ}$ increments. A stationary random dot ``null'' stimulus was interleaved with the motion stimuli in 36 of 46 recorded neurons. For these 36 neurons, stimuli were repeated 51-223 times. The remaining 10 neurons had 20-30 stimulus repetitions. In all experiments, dot textures appeared and remained stationary for 256 ms, translated for 256 ms with constant direction and speed, and were again stationary for 256 ms. When the motion of a dot carried it outside of the aperture, it was randomly positioned along the leading aperture edge to maintain dot number. Trials were separated by a brief pause of 1-2s. Spike waveforms were sampled at 10kHz and isolation was aided by a window discriminator. Spike times were determined by threshold crossings. Data from three of the four monkeys have been previously published \cite{osborne_time_2004,osborne_neural_2008}. 

Similar ``alert motion-step'' experiments were performed with two adult male monkeys (\textit{Macaca mulatta}) that maintained fixation during visual stimulus presentation. These methods have been described in more detail elsewhere \cite{liu_efficient_2016,mukherjee_2017}. Animals were pair-housed when possible and had daily access to enrichment activities and play areas. Pre-study instrumentation with a head stabilization post, an eye coil, and a recording chamber was performed under isoflurane anesthesia using sterile surgical technique and post-operative analgesia with buprenorphine. Animals were trained over a period of time to acclimate them to the laboratory environment. Daily experiments involved seating the monkey in a plastic ``chair'' in front of a visual display in a dimly lit room. The task required fixation within $2^\circ$ throughout the 2-3 second trial to obtain a juice reward. Eye position was monitored via a surgically implanted scleral coil. We employed similar visual stimuli to the anesthestized experiments. We presented bright random dot stimuli against the dark screen of a CRT display set to 1024x768 resolution and a 100 Hz frame rate (Sony GWFM-FW9011). Recordings were made with 3 quartz-platinum/tungsten single microelectrodes (TREC, Germany). We sampled the voltage waveforms from the array at 30 kHz (Plexon Omniplex) and stored them for offline analysis. As in the anesthetized experiments, we performed online analyses to map the direction and speed tuning, and the size and location of each unit's excitatory receptive field. Eight of the thirty-four neurons were recorded as simultaneous pairs, and thus did not have perfectly optimized stimulus speed and size for both cells in the pair. We confirmed unit isolation through principal component analysis of spike waveforms along with inspection of interspike interval distributions. Motion stimuli were comprised of 24 directions, evenly spread between the preferred direction of the isolated single unit and $\pm180^\circ$. A stationary random dot ``null'' stimulus was interleaved with the motion stimuli.

\subsection*{Analytical methods}

We computed the spike count in the 250 ms following motion onset, while the dot textures were translating. We computed the Fano factor of the spike count (variance divided by mean) as a function of stimulus direction using a repeated-measures ANOVA on the null hypothesis of a constant Fano factor across directions. Stimulus direction relative to the preferred direction of the isolated single unit was found to have an effect on Fano factor in the alert experiments $(p=7.4\times10^{-7})$ but not in the anesthetized experiments $(p=0.42)$. Because there was a narrower range of stimulus directions in the anesthetized experiments, we limited both data sets to stimulus directions within $\pm90^{\circ}$ of the preferred direction.

Tuning widths were obtained by fitting a Gaussian curve to the spike counts as a function of direction for each neuron. The tuning width was taken to be the standard deviation of the Gaussian of best-fit. Direction selectivity was computed as a direction index (DI), where
\begin{equation}
\mathrm{DI}=\frac{r_\mathrm{pref}-r_\mathrm{orth}}{r_\mathrm{pref}+r_\mathrm{orth}}.
\label{eq:direction_index}
\end{equation}
Here, $r_\mathrm{pref}$ is the mean response to the preferred stimulus direction and $r_\mathrm{orth}$ is the mean response to the orthogonal directions.
Similarly, a variance tuning index was calculated as
\begin{equation}
\mathrm{VTI}=\frac{\sigma^2_\mathrm{pref}-\sigma^2_\mathrm{orth}}{\sigma^2_\mathrm{pref}+\sigma^2_\mathrm{orth}}.
\label{eq:variance_index}
\end{equation}
FFTI was defined analogously.
\begin{equation}
\mathrm{FFTI}=\frac{\mathrm{FF}_\mathrm{orth}-\mathrm{FF}_\mathrm{pref}}{\mathrm{FF}_\mathrm{orth}+\mathrm{FF}_\mathrm{pref}}.
\label{eq:fanoFacor_index}
\end{equation}

The procedure for mean-matched Fano factor is adapted from \cite{churchland_stimulus_2010}. The distribution of mean spike counts was computed for each condition. The distribution of spike counts for each stimulus direction was approximated by binning values in 15 evenly spaced bins that spanned the range of responses. For each bin in which one condition had more data points than the other, data points were randomly discarded from that condition until both distributions matched. The Fano factor was calculated for each neuron in that stimulus condition. The mean Fano factor for each stimulus condition was used to calculate FFTI for each subject condition. This resampling procedure was repeated one million times to generate a distribution of Fano factor tunings. The resampled Fano factor values were fit to cosine curves by a least squares fitting procedure.

The Fisher information was calculated for simulated neuronal populations using
\begin{equation}
J(\theta) \approx J_{\textrm{mean}}(\theta) = \vec{f}'(\theta)^T \Sigma^{-1}(\theta) \vec{f}'(\theta),
\label{eq:fisher_information}
\end{equation}
where the derivative is taken with respect to motion direction $\theta$, and $\vec{f}$ is a vector containing the tuning curves for each neuron in the population, and we ignore the contributions to the Fisher information from derivatives of the covariance matrix \cite{ecker_2011,zylberberg_direction-selective_2016}. The covariance matrix between neurons is given by $\Sigma(\theta)$. The diagonal elements contain each neuron's variance, as imposed by the Fano factor tuning curves we model in the three qualitative regimes (U-shaped tuning of the variance relative to the mean, flat, and inverted-U). The off-diagonal elements of $\Sigma(\theta)$ are imposed by our correlation model, which is a von Mises distribution, as given by Eq \ref{corr_equation}. For the homogeneous population, each neuron has the same tuning curve, given by a von Mises distribution 
\begin{equation}
f(\theta) = b + A\frac{e^{\kappa(cos(\theta-\theta_{\textrm{pref}})+1)}-1}{e^{2\kappa}-1}
\end{equation}
fit to the population averaged tuning in each behavioral state, where $b$ and $A$ are background firing rates and peak firing rates, respectively; $\theta_{\textrm{pref}}$ is the preferred direction of the neuron; $\theta$ is the stimulus direction; and $\kappa$ is the width of the tuning curve. Preferred directions in the model population are distributed evenly across all recorded directions. For the heterogeneous population model, tuning curves are again modeled with von Mises distributions fit to the diversity of tunings measured in each behavioral state. Populations models are built up by sampling with replacement from the recorded, fit tuning curves. 

Mutual information was calculated for each neuron from cumulative spike counts across 13 stimulus directions, in 15 degree increments $\pm90^{\circ}$ around the preferred direction. The mutual information, $I$, between spike count $k$ and stimulus direction $\theta$ is given by
\begin{equation}
I(k;\theta)=\sum_{\theta} p(\theta) \sum_k p(k|\theta)\textrm{log}_2\frac{p(k|\theta)}{p(k)}.
\end{equation}
Sampling bias in information estimates were corrected via bootstrap resampling at fractions of the data between $95-50\%$; finite size effects were estimated using the method of quadratic extrapolation \cite{panzeri_2007}. 


\nolinenumbers

\clearpage

\section*{Supporting Information}
\renewcommand{\thefigure}{S\arabic{figure}}
\setcounter{figure}{0}  

\begin{figure}[ht]
	\centering
		\includegraphics[width=0.75\textwidth]{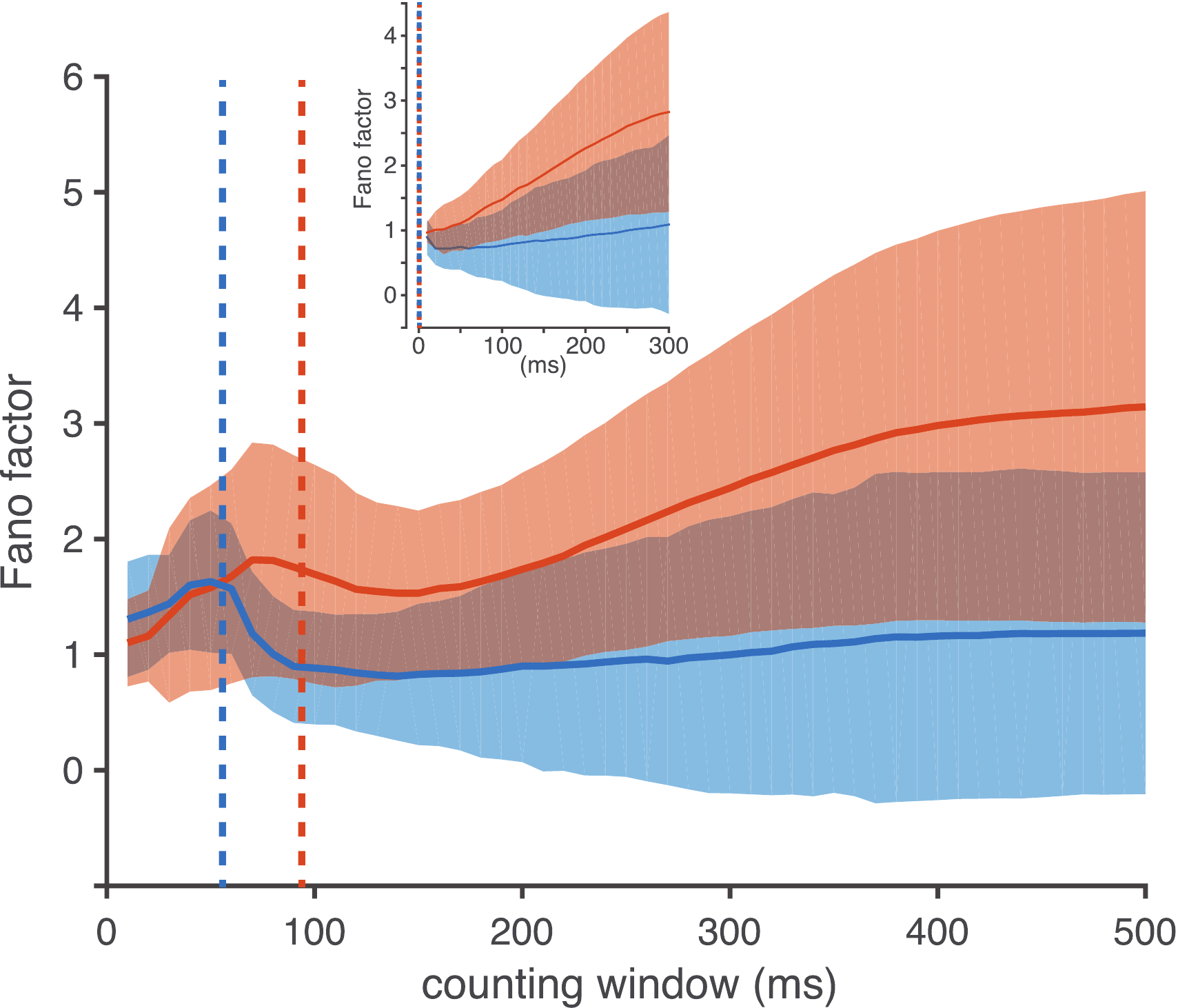}
		\caption{Fano factors over expanding time windows in the alert state (blue) and the anesthetized state (orange). Solid traces indicate population means and shaded areas indicate the standard deviation across the each neural population. Time windows begin at stimulus motion onset. Stimulus motion is in the preferred direction for each neuron. The average response latency for each population is indicated by the vertical dashed lines at 56ms and 94ms, for the alert and anesthetized data, respectively. Inset: Spike count windows are aligned with response onset. Fano factors over expanding time windows in the alert state (blue) and the anesthetized state (orange) aligned by each neuron's response onset time. }
	\label{fig:ff_window_fig}
\end{figure}

\begin{figure}
	\includegraphics[width=1\textwidth]{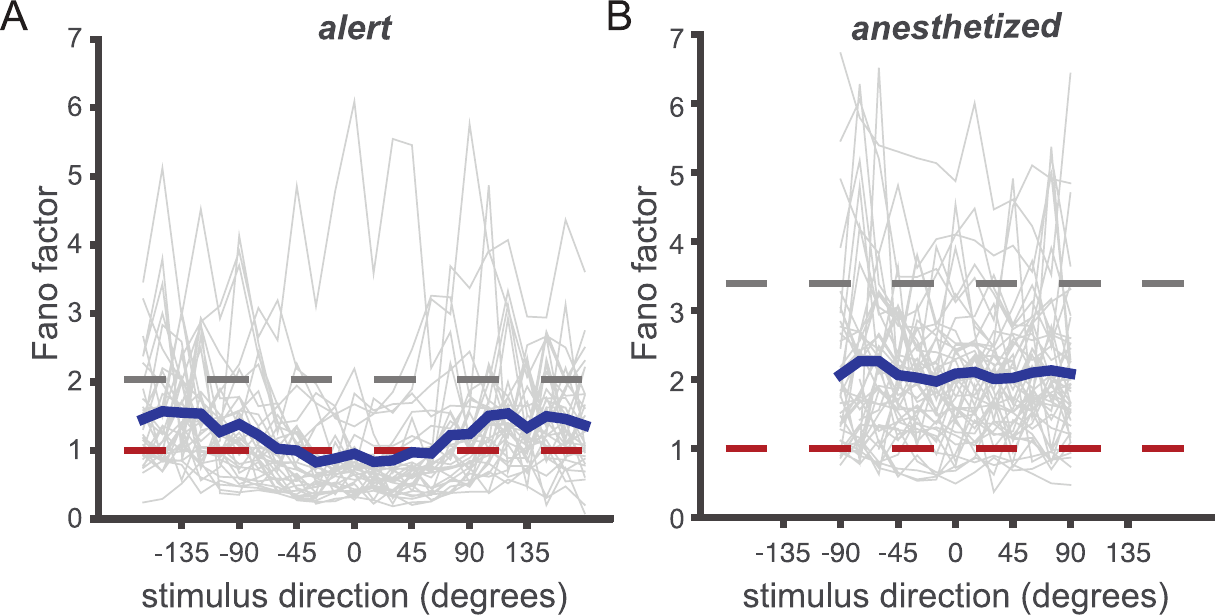}
		\caption[Mean Fano factor by stimulus direction exhibits the same state-dependence of stimulus-induced variability as median]{Mean Fano factor by stimulus direction exhibits the same state-dependence of stimulus-induced variability as median. Same as Fig 3C-D but with the mean Fano factor for each stimulus direction shown in blue rather than the median.}
	\label{fig:mean_fano}
\end{figure}

\begin{figure}[ht]
	\centering
		\includegraphics[width=1\textwidth]{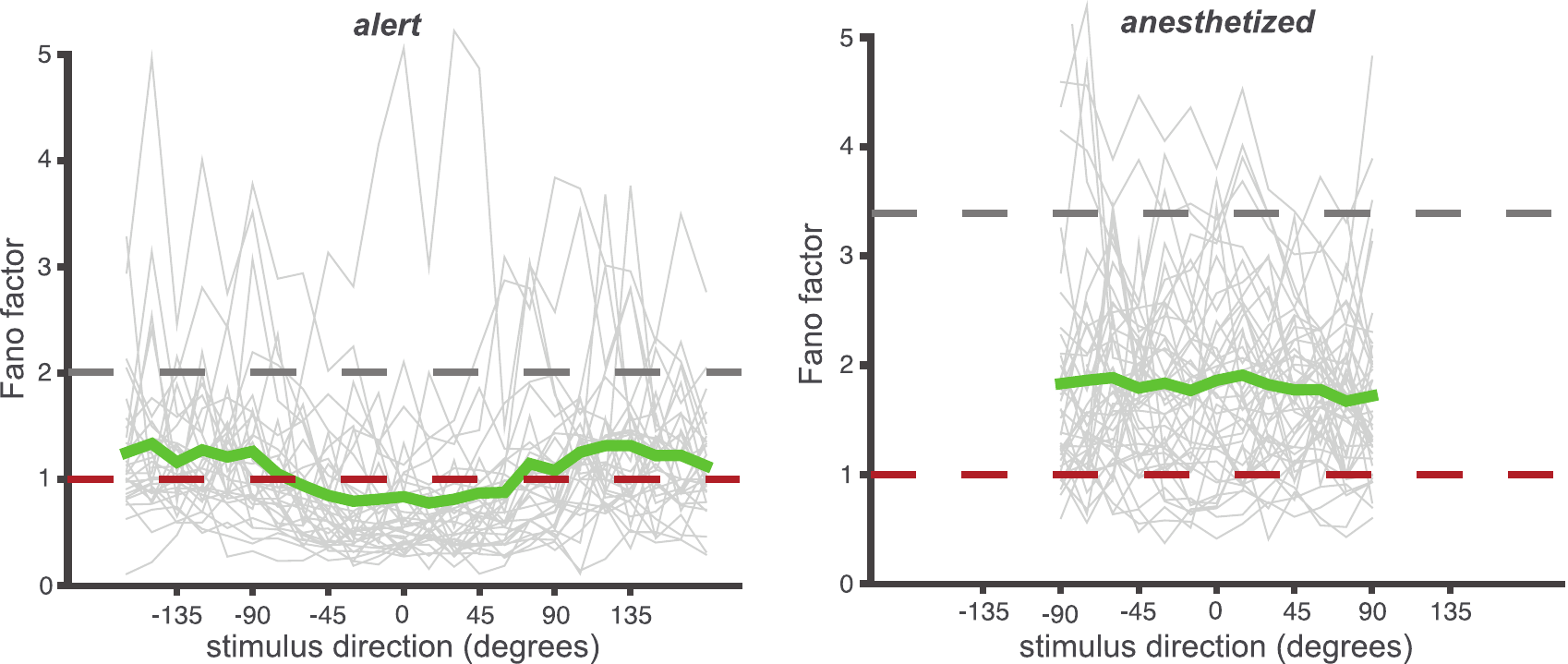}
		\caption[Aligning spike count windows by response onset does not affect stimulus-dependent Fano factor tuning]{Aligning spike count windows by response onset does not affect stimulus-dependent Fano factor tuning. Figure is as in Fig 3C-D but spike count windows are aligned to response onset rather than stimulus motion onset. Neurons in the alert state tend to have shorter latencies than those in the anesthetized state, but this does not affect their mean Fano factor or its tuning in either state.}
	\label{fig:latency_aligned_fano}
\end{figure}

\begin{figure}[ht]
	\centering
		\includegraphics[width=0.5\textwidth]{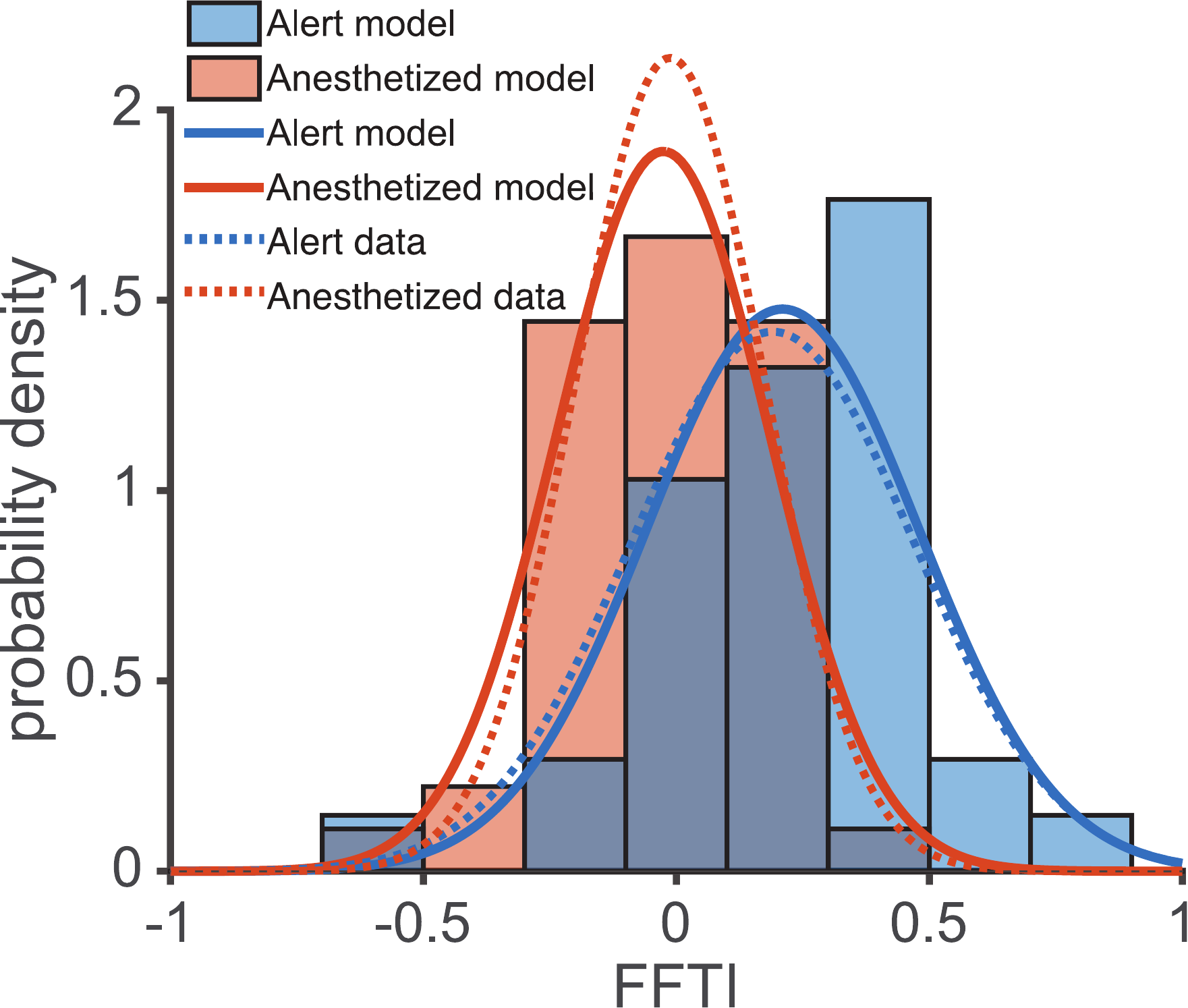}
		\caption[Aligning spike count windows by response onset preserves the distributions of Fano factor tunings]{Aligning spike count windows by response onset preserves the distributions of Fano factor tunings and the shift to larger, positive tuning indices in the alert state. Figure is as in Fig 3E but spike count windows are aligned to response onset rather than stimulus motion onset. Blue and orange traces are Gaussian best fits to FFTI distributions in alert and anesthetized states, respectively. The dashed traces are Gaussian fits to the FFTI distributions to the data.}
	\label{fig:latency_aligned_ffti}
\end{figure}

\begin{figure}
	\includegraphics[width=1\textwidth]{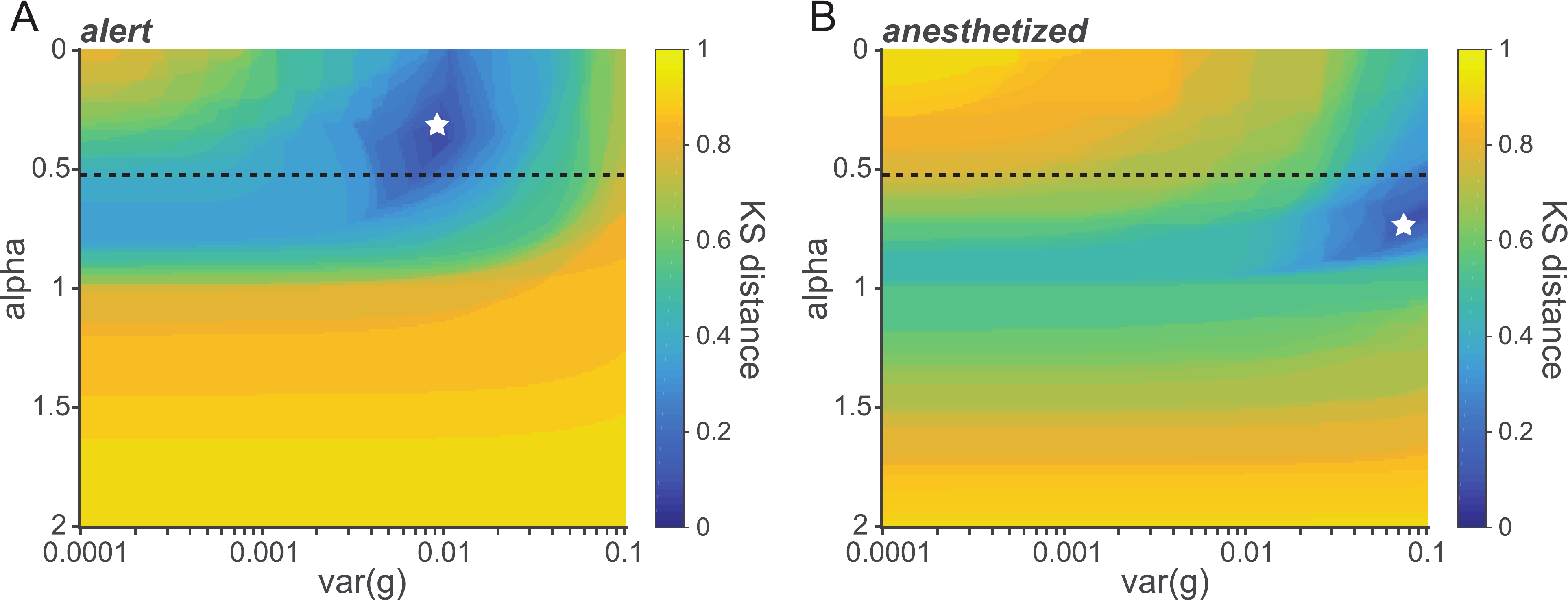}
		\caption[]{Heat maps illustrate the quality of fit of the variance model for the alert (A) and anesthetized (B) states. In this model, a single value of $\alpha$ and var(g) is fit for each population. The parameter values are applied via the variance model to the tuning curves for each population, which returns a distribution of FFTI values. Quality of fit is measured by minimizing the Kolmogorov-Smirnov distance between the model FFTI distribution and the observed distribution. The KS test statistic is shown for a range of parameter values, alpha and var(g), for each population. The optimal parameter values used in Fig 6A are indicated with white stars. The optimal parameters for the alert state are $\alpha=0.31$ and $\textrm{var(g)}=0.0094$. The optimal parameters for the anesthetized state are $\alpha=0.74$ and $\textrm{var(g)}=0.0732$. The dashed lines indicate a best compromise $\alpha$ parameter found by minimizing the mean-squared KS statistic for both states. This value, $\alpha = 0.53$, was used in the model in Fig \ref{fig:common_alpha_ffti}.}
	\label{fig:model_fitting_ks}
\end{figure}

\begin{figure}
	\centering
	\includegraphics[width=0.5\textwidth]{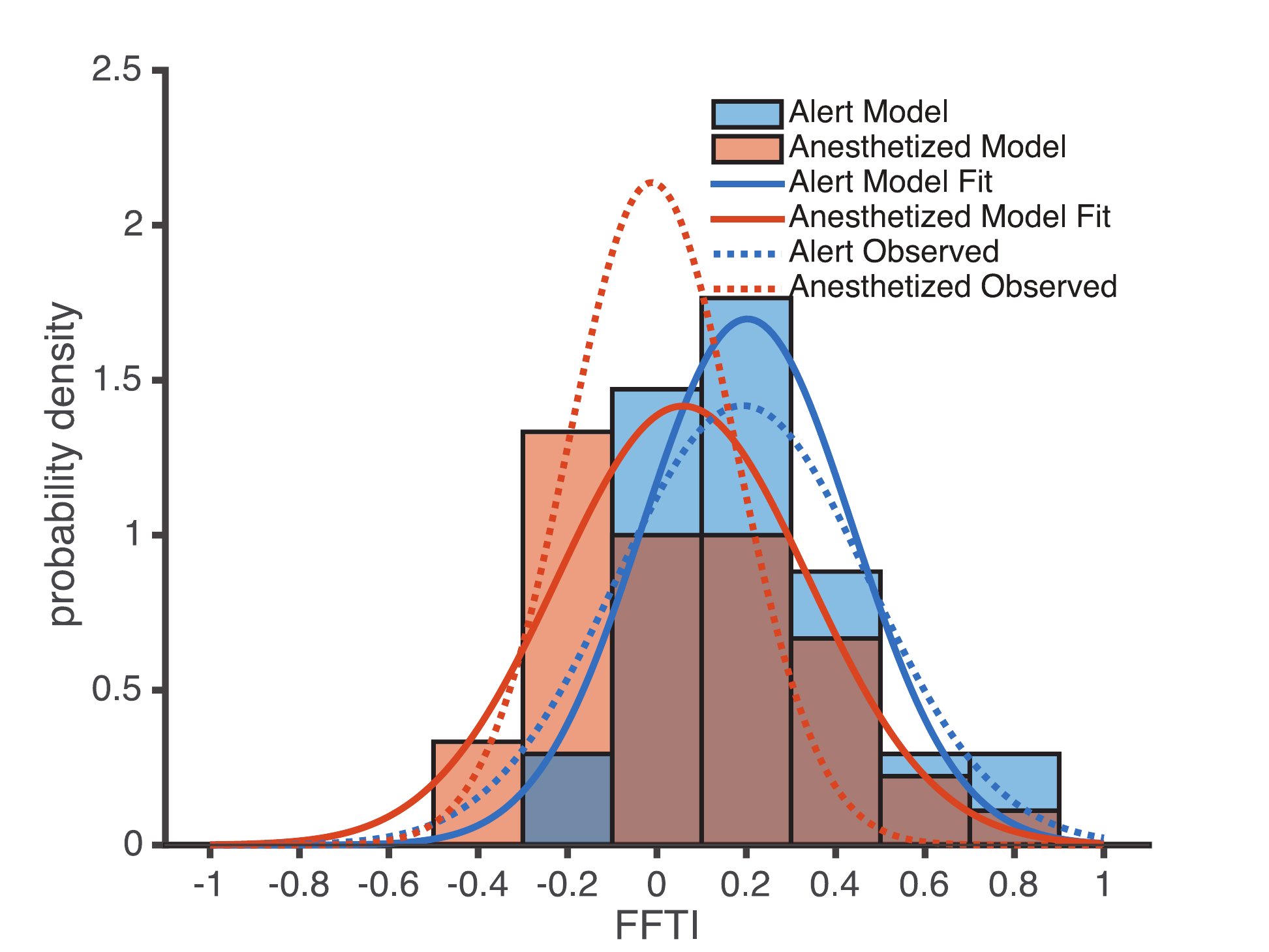}
		\caption[]{Fitting the FFTI distributions with a common value of $\alpha$ in each population. This model is similar to Figure 6A, but with a single compromise value of $\alpha=0.53$ shared between each population and separate values of $\textrm{var(g)}=0.0079$ in the alert state and $\textrm{var(g)}=0.1000$ in the anesthetized state. The model is still able to capture most of the distribution of FFTI observed in the real populations. The common value of $\alpha$ was determined by finding the value of $\alpha$ that minimized the sum of the squared KS distance for each population, with var(g) allowed to vary freely. The ability of this model to capture most of the differences in FFTI distributions between states suggests that changes in var(g) could be the primary cause of difference in Fano factor tuning between states..}
	\label{fig:common_alpha_ffti}
\end{figure}

\begin{figure}
	\centering
	\includegraphics[width=0.5\textwidth]{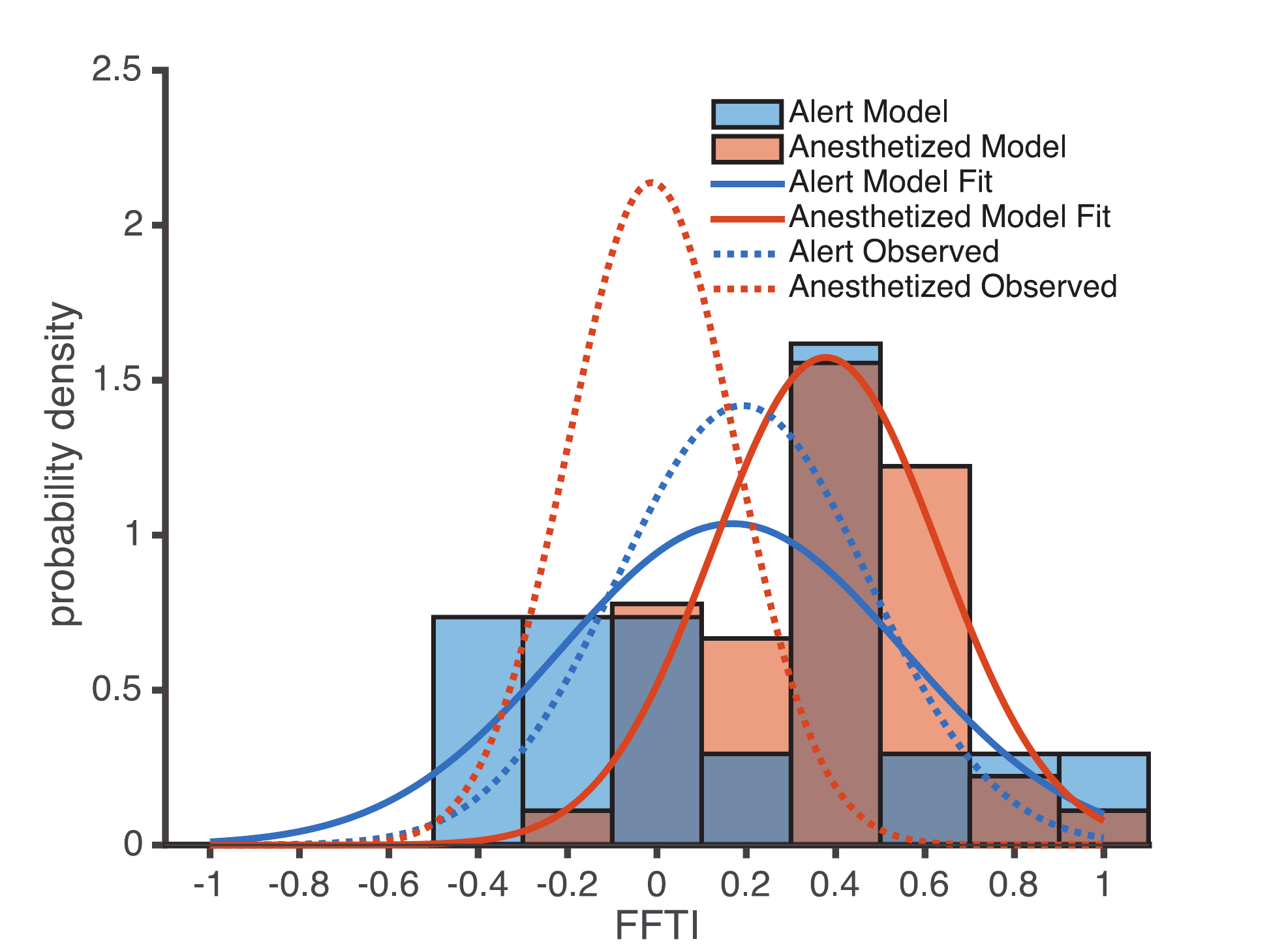}
		\caption[]{Fitting the FFTI distributions with a common value of var(g) in each population. This model is similar to Figure 6A and supplementary Figure \ref{fig:common_alpha_ffti} , but with a single compromise value of var(g)=0.0105 shared between each population and separate values of $\alpha=0.3216$ in the alert state and $\alpha=0.4945$ in the anesthetized state. Unlike the the model in Fig \ref{fig:common_alpha_ffti}, the different distributions of FFTI between states cannot be explained by changes in the $\alpha$ parameter alone. In fact, this model has the opposite qualitative shift in the FFTI between states: the anesthetized state now has more tuning in the Fano factor than the alert state.}
	\label{fig:common_varg_ffti}
\end{figure}

\begin{figure}
	\centering
	\includegraphics[height=.7\textheight]{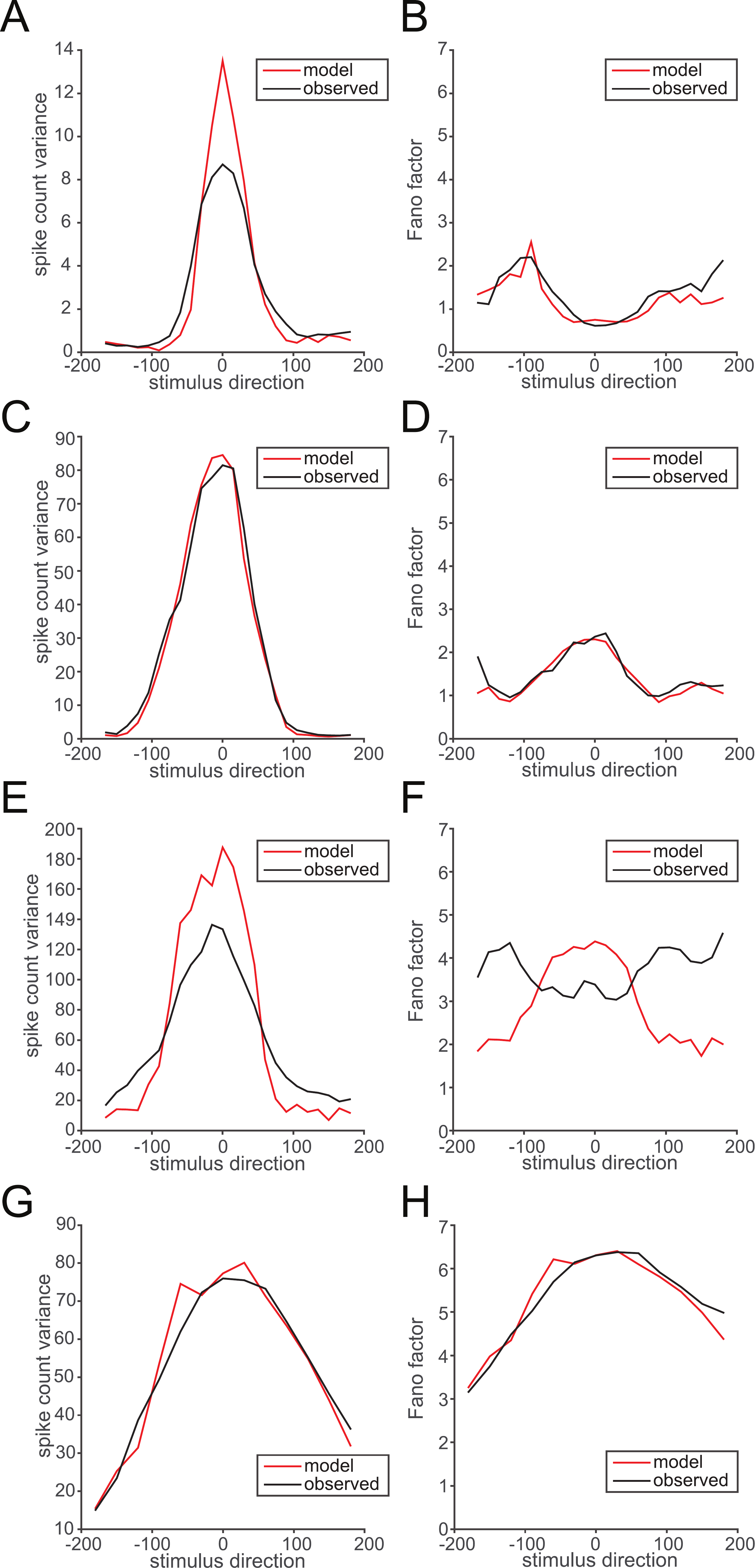}
		\caption[]{The variance model fits Fano factor across stimulus direction for example neurons. The variance model is able to fit a variety of Fano factor tunings. The model used a least squares fitting to find the optimal $\alpha$ and var(g) parameters for each neuron to match the observed Fano factor across stimuli. (A,C,E,G) The observed spike count variance (black trace) and model fit (red trace) for four example neurons across stimulus directions. (B,D,F,H) The observed Fano factor (black trace) and model fit (red trace) for the same example neurons in (A,C,E,G). Representative examples were chosen to demonstrate the variety of Fano factor tunings that the model is able to represent. The model fit in E and F shows the limitations of the model, which fails to fit the Fano factor tuning for some neurons. The model tends to fit poorly for neurons with a positive FFTI and a large Fano factor, as increasing the Fano factor in the model, by increasing $\alpha$ or var(g), will tend to decrease the FFTI.}
	\label{fig:sample_fits}
\end{figure}

\begin{figure}
	\centering
	\includegraphics[width=0.5\textwidth]{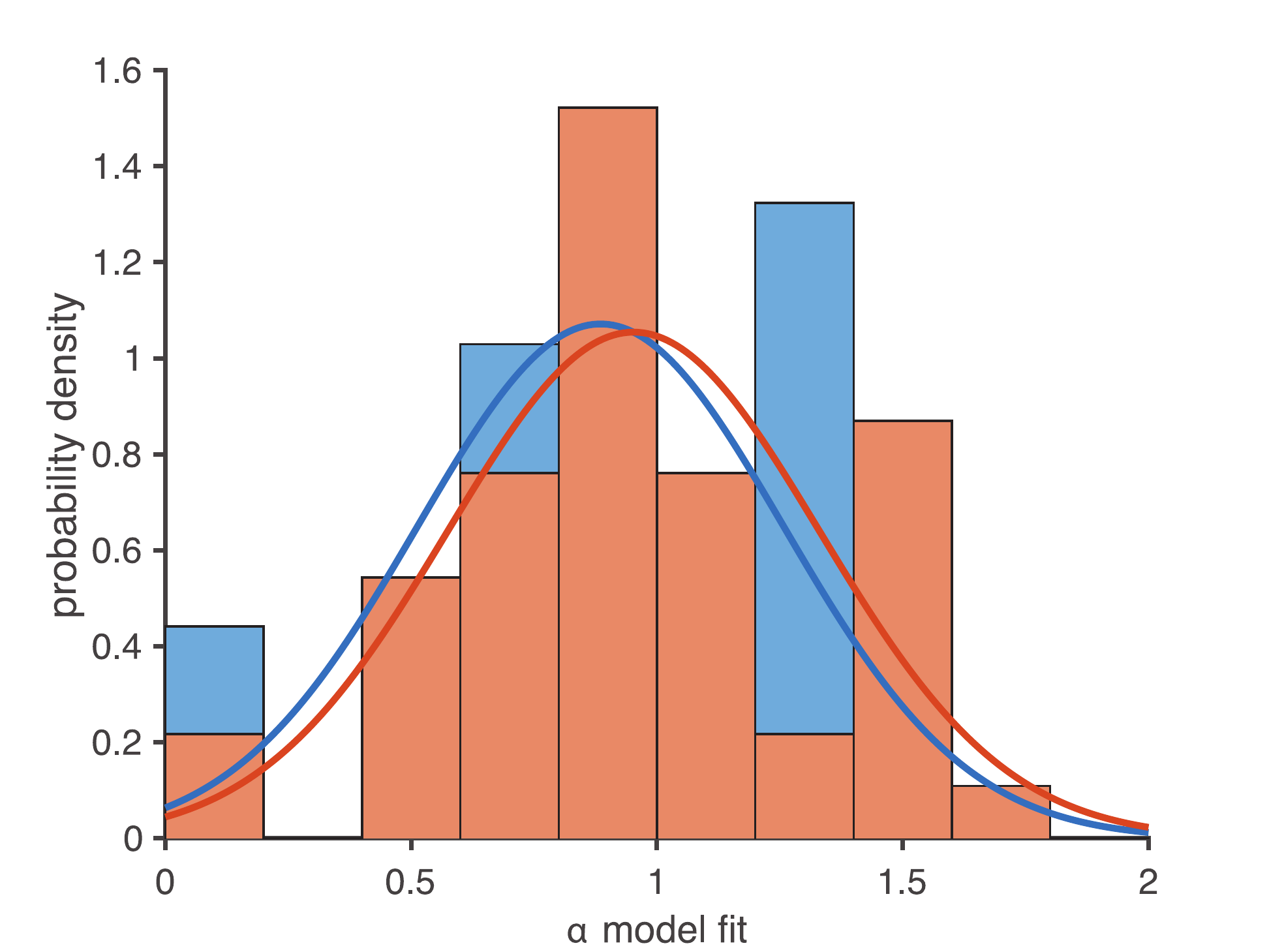}
		\caption[]{The distribution of $\alpha$ parameters fit in the model across all neurons. The model fits are the same as shown in Fig \ref{fig:model_fitting_ks} of the supplementary materials, where parameters are fit for each neuron to match the Fano factor across stimuli given the mean responses. The blue trace (alert) and orange trace (anesthetized) show the Gaussian best fit to the distribution. The mean values of $\alpha$ for each population are 0.89 for the alert state and 0.95 for the anesthetized state.}
	\label{fig:alpha_fits}
\end{figure}

\begin{figure}
	\centering
	\includegraphics[width=0.5\textwidth]{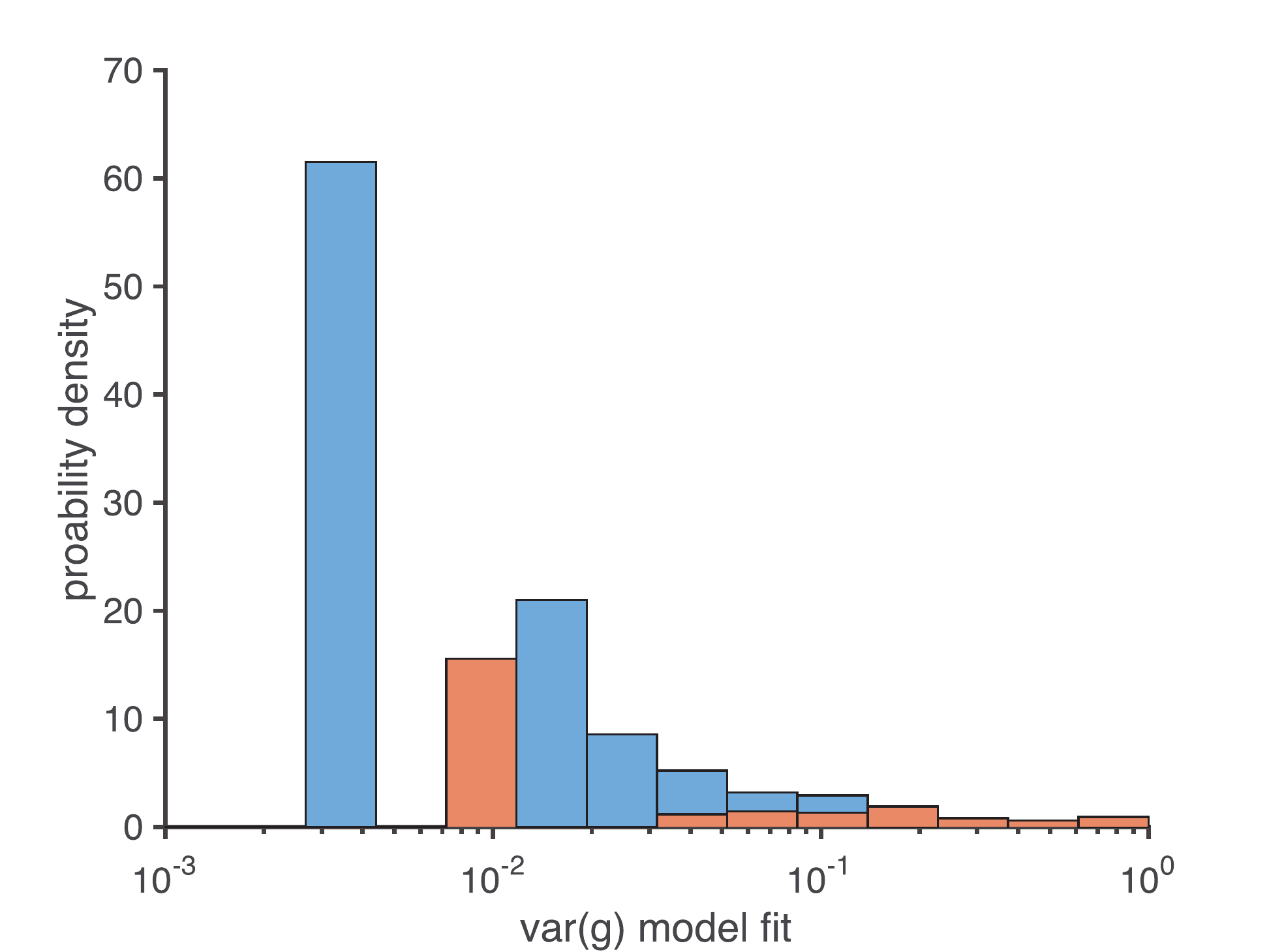}
		\caption[]{The distribution of var(g) parameters fit in the model across all neurons. The model fits are the same as shown in Figs \ref{fig:model_fitting_ks} and \ref{fig:alpha_fits}, where parameters are fit for each neuron to match the Fano factor across stimuli given the mean responses. The histograms show the distribution of parameters for the alert (blue) and anesthetized (orange) states. The values of var(g) fit to the anesthetized data are greater than the values fit to the alert state. The mean values of var(g) for each population are 0.062 for the alert state and 0.43 for the anesthetized state, again showing a nearly order-of-magnitude change in this parameter.}
	\label{fig:varg_fits}
\end{figure}

\begin{figure}[ht]
	\centering
		\includegraphics[width=1\textwidth]{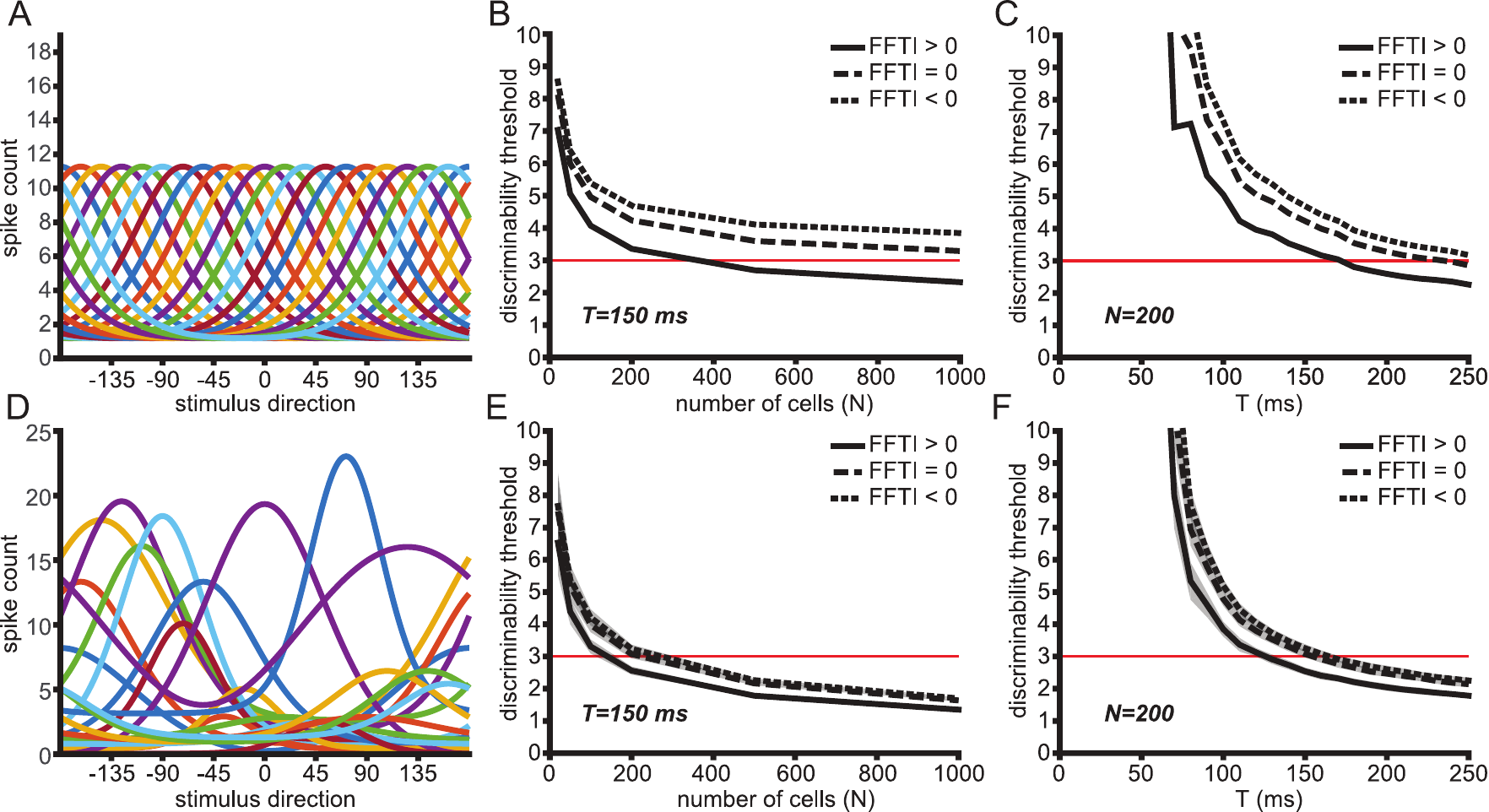}
		\caption[Information-limiting correlations do not affect relative decoding performance between different variance models]{Fano factor tuning and heterogeneity both contribute to lower discriminability thresholds in MT populations. Similar to Fig 7, but without information-limiting correlations. (A) Model population with 20 homogeneous tuning curves. (B) Cramer-Rao bound in homogeneous population models of different sizes with short-range correlations ($c_{max}=0.1$, $\langle c\rangle=0.0438$). Tuning curves are fit to the average cumulative response up to 150 ms after motion onset. Black traces show performance of models with varying stimulus-dependent variance. The solid trace is $\textrm{FFTI}>0$, the dashed trace is $\textrm{FFTI}=0$, and the dotted trace is $\textrm{FFTI}<0$. The red line indicates the stimulus discriminability threshold for smooth pursuit behavior in macaques 125 ms after pursuit initiation. (C) Same as in (B) but for populations of 200 neurons with first-order statistics matched to average response at time $t$ after motion onset. (D) Sample population of 20 heterogeneous tuning curves drawn from measured tuning curves in recorded neurons. (E,F) Same as in (B,C) but for heterogeneous populations. The shaded areas show the standard deviation of the Cramer-Rao bound.}
	\label{fig:model_wo_limited_info}
\end{figure}

\end{document}